\documentclass[fleqn,10pt]{wlscirep}
\usepackage[utf8]{inputenc}
\usepackage[T1]{fontenc}
\usepackage{authblk}
\usepackage{amssymb}
\usepackage{amsmath} 
\usepackage{amsfonts}
\usepackage{latexsym}
\usepackage{amsthm} 
\usepackage{eucal}

\usepackage{rotating}
\usepackage[ruled,vlined,resetcount,algosection,boxed,algo2e]{algorithm2e}
\DontPrintSemicolon  
\usepackage{subfig}
\usepackage{graphicx}
\usepackage[english]{babel}
\usepackage[english]{varioref}
\usepackage{relsize}%
\usepackage{url}
\usepackage{multirow}
\usepackage{hhline}
\usepackage{colortbl}
\usepackage{wrapfig}
\usepackage{microtype}
\usepackage{tikz}
\usepackage{pgfplots}
\usepackage{caption}
\usepackage[subfigure]{tocloft}
\pgfplotsset{compat=1.18} 

\usepackage{xfrac}
\usepackage{enumerate}
\usepackage[overload]{empheq}
\usepackage{mathtools}
\usepackage{cases} 

\newcommand{\SI}[1]{\textcolor{blue}{{\it Supplementary Information}, #1}}

\DeclareMathOperator*{\argmin}{arg\,min}

\addto\captionsenglish{
  \renewcommand{\contentsname}%
    {Supplemental Information:\\ The temporal dynamics of group interactions in higher-order social networks}%
}

\newcommand{\listequationsname}{Supplementary Equations}
\newlistof{myequations}{equ}{\listequationsname}

\addto\captionsenglish{%
  }
\def\l@section{\@tocline{1}{0,2pt}{2pc}{12mm}{\ \ }} 

\newcommand{\startsupplement}{
    \setcounter{figure}{0}
    \setcounter{table}{0}
    \setcounter{equation}{0}
    \renewcommand{\thetable}{S\arabic{table}}
    \renewcommand{\theequation}{S\arabic{equation}}
    \renewcommand{\thefigure}{S\arabic{figure}}
    \renewcommand{\cftfigpresnum}{Fig.~}
    \renewcommand{\cfttabpresnum}{Tab.~}
    \setlength{\cftfignumwidth}{5.5em}
    \setlength{\cftfigindent}{0em}
    \setlength{\cfttabnumwidth}{5.5em}
    \setlength{\cfttabindent}{0em}
}

\title{The temporal dynamics of group interactions in higher-order social networks}

\makeatletter

\author[1,2,3,*]{Iacopo Iacopini}
\author[3,4]{M\'arton Karsai}
\author[5]{Alain Barrat}

\affil[1]{Network Science Institute, Northeastern University London, London, E1W 1LP, United Kingdom}
\affil[2]{Department of Physics, Northeastern University, Boston, MA 02115, USA}
\affil[3]{Department of Network and Data Science, Central European University, A-1100 Vienna, Austria}
\affil[4]{National Laboratory for Health Security, HUN-REN Alfr\'ed R\'enyi Institute of Mathematics, 1053 Budapest, Hungary}
\affil[5]{Aix Marseille Univ, Universit\'e de Toulon, CNRS, CPT, Turing Center for Living Systems, Marseille, France}

\affil[*]{corresponding author: iacopo.iacopini@nulondon.ac.uk}

\begin{abstract}
Representing social systems as networks, starting from the interactions between individuals, sheds light on the mechanisms governing their dynamics. However, networks encode only pairwise interactions, while most social interactions occur among groups of individuals, requiring higher-order network representations. Despite the recent interest in higher-order networks, little is known about the mechanisms that govern the formation and evolution of groups, and how people move between groups. Here, we leverage empirical data on social interactions among children and university students to study their temporal dynamics at both individual and group levels, characterising how individuals navigate groups and how groups form and disaggregate. We find robust patterns across contexts and propose a dynamical model that closely reproduces empirical observations. These results represent a further step in understanding social systems, and open up research directions to study the impact of group dynamics on dynamical processes that evolve on top of them.
\end{abstract}

\begin{document}

\flushbottom
\maketitle
\thispagestyle{empty}

\section*{Introduction}
Social interactions are the building blocks of our society~\cite{wasserman1994social}. 
Humans ---and animals in general--- form groups of different sizes~\cite{lehmann2007group} and have learnt the advantages of communicating and gathering in close social circles~\cite{dunbar2018anatomy, dunbar2020structure}. 
Our everyday social path is in fact a succession of these group events that involve different numbers of peers, from walking alone or having a coffee with a friend, to engaging in meetings or group conversations at work or during social gatherings. Networks provide a powerful tool to represent these complex social trajectories with the capacity to encode the structure and dynamics of interactions between individuals~\cite{albert2002statistical, newman2003structure,barrat2008dynamical,latora_nicosia_russo_2017}. 
The use of network representations and of social network analysis tools~\cite{wasserman1994social}, as well as the emerging field of temporal networks, have helped identifying the mechanisms that govern the formation and evolution of these structures~\cite{vespignani2018twenty, holme2012temporal, holme2015modern}.
Nevertheless, these conventional network descriptions are inherently limited to the description of pairwise interactions, which does not capture the full complexity of the social phenomena~\cite{battiston2020networks, battiston2021physics, torres2021and, bianconi_2021}. Considering interactions of higher-order is thus compulsory to represent and model how humans interact in groups~\cite{milojevic2014principles, juul2024hypergraph} or how animals gather~\cite{katz2011inferring}.

The structure and dynamics of group interactions, however, are complex~\cite{mcgrath1984groups}. Groups may have heterogeneous sizes~\cite{patania2017shape, benson2018simplicial, cencetti2021temporal, iacopini2022group, korbel2023homophily}, can  change dynamically~\cite{forsyth2018group, geard2010competition} or exhibit hierarchical and nested structures~\cite{lotito2022higher, mancastroppa2023hyper}. 
Possible driving mechanisms behind these characters include for example, simplicial closure~\cite{benson2018simplicial} and homophily~\cite{korbel2023homophily}.
Most studies on group formation and structure, however, do not take into account the further temporal evolution of the underlying social systems ---that is characterised by patterns of memory and burstiness, and exhibits a complex dynamics of merging and splitting of groups~\cite{zhao2011social,cencetti2021temporal,ceria2023temporal, gallo2024higher}.
For instance, larger groups tend to have shorter durations \cite{zhao2011social} and exhibit shorter temporal correlations \cite{gallo2024higher}; the dynamics of group formation and fragmentation exhibits a preferred temporal direction \cite{cencetti2021temporal,gallo2024higher}, and non-trivial recurrence of groups can emerge, driven by different contexts and geographical places of interactions and defining social circles~\cite{sekara2016fundamental}.
These complex patterns are the results of microscopic individual level decisions, ultimately shaping the emergence of collective behaviours. Understanding these mechanisms is essential to better characterise the emerging group dynamics and their effects on processes such as disease transmission or spread of information, social norms and behavioural patterns within and across group gatherings~\cite{barrat2008dynamical, castellano2009statistical, vespignani2012modelling, pastor2015epidemic}.\\

Here, we address this challenge by investigating empirical traces of group dynamics extracted from proximity data in different social and temporal contexts. Leveraging two data sets of temporally-resolved human interactions among preschool children and freshmen students, we highlight complex mechanisms of group dynamics both at the individual and at the group level. 
Following the group membership of individuals across groups of different sizes, we find that the main dynamical patterns of group-change are independent from the context of interactions. The statistics of group durations exhibit as well robust properties, with a {\it long-gets-longer} effect \cite{zhao2011social,vestergaard2014memory} for all group sizes (i.e., the probability to change group decreases with the time spent in it). Furthermore, the dynamics of group aggregation and disaggregation show hierarchical and largely symmetrical properties of assembly and disassembly.
Finally, we propose a dynamical model for temporal interactions ---that takes groups explicitly into account--- and show that it reproduces the empirical patterns.

Our results shed lights on how temporal patterns of group formation and evolution can result from microscopic choices at the level of individuals, by accounting for mechanisms of social and temporal memory. The proposed model can moreover serve as a synthetic structure for studying the impact of group interactions and their temporal properties on dynamical processes: indeed, recent works based on static hypergraphs have shown evidence that group interactions can induce critical mass effects in social contagion~\cite{iacopini2019simplicial, st2022influential}, amplify small initial opinion biases and accelerate the formation of consensus~\cite{papanikolaou2022consensus, iacopini2022group} and cooperation~\cite{sheng2024strategy, civilini2024explosive}, but investigations on evolving structures are scarce \cite{Chowdhary_2021}. Overall, our study provides a starting point for increasingly realistic modelling approaches to better characterise complex social systems 
and the phenomenology of attached processes.

\section*{Results}

\subsection*{Extracting groups from real-world data}

We consider records from two data-collection efforts that tracked social interactions at a university and in a preschool, yielding data sets in the form of temporal networks, in which each person is represented as a node and each interaction as a temporal edge (see \textit{Methods}). Time is in each case discretized by the temporal resolution of the data collection setup. At each timestamp, we define as groups the maximal cliques (largest fully connected subgraphs) and build in this way a temporal hypergraph. At each timestamp, each node can thus be either isolated or part of one or several groups (hyperedges).

\paragraph*{University.} We use data collected by the Copenhagen Network Study (CNS)~\cite{sapiezynski2019interaction}. It is a temporally-resolved data describing the proximity events of 706 freshmen students at the Technical University of Denmark, collected using the exchange of Bluetooth signals by their smartphones.
We use the publicly available data describing these proximity events during four consecutive weeks and with a temporal resolution of $5$ minutes \cite{sapiezynski2019interaction}.
We split the data into three different contexts, which might result in different interaction patterns. First, we treat all interactions taking place over the {\it weekends} as a separate set. Second, we divide interactions that happen during the workweek into {\it in-class} and {\it out-of-class} time. 
In this way, we do not mix the group dynamics emerging in unconstrained interactions during the free time of the lunch break and in-between classes with potentially constrained co-presence due to common attendance of classes and/or seating configurations. For this data set moreover, we perform some data pre-processing before extracting the groups in each timestamp, to filter out very weak interactions (based on the Bluetooth Received Signal Strength Indication), smoothen intermittent patterns, remove spurious connections, and perform a standard triadic closure procedure with tailored parameters~\cite{sekara2016fundamental}. Additional details on the data set and the pre-processing are described in the {\it Methods}. 

\paragraph*{Preschool.} We also consider another data set, collected in a preschool as part of the DyLNet project~\cite{dai2022longitudinal} to follow the social interactions of children of age $3-6$ and their teachers and assistants. 
The data describes proximity social interactions between $174$ children and $34$ adults in seven classes, recorded by Radio Frequency Identification (RFID) Wireless Proximity Sensors carried by each participant. Interactions were recorded with a temporal resolution of $5$ seconds, during periods of $5$ consecutive days, for $10$ consecutive months (overall $50$ days of data collection) of a single academic year in a French pre-school. For the purpose of our study we rely on a pre-processed data set shared in~\cite{dai2022longitudinal} and a temporal network reconstructed from the cleaned interaction signals as explained in~\cite{dai2020temporal}, and we remove the data of interactions with and between adults, to focus on the childrens' group dynamics.

Similarly to the university setting, we also divide the data according to the contexts that may impose different constraints on the emergence of possible group interactions. We differentiate between {\it in-class} periods, during which the social grouping of children was strongly influenced by the teachers' instructions and scheduled activities, and {\it out-of-class} periods, when children could choose freely to interact with anyone from their own and potentially other classes. For more details on the data pre-processing, network reconstruction, and context selection, we refer to the {\it Methods} section.


\subsection*{The dynamics of group change}

\begin{figure*}
	\centering
	\includegraphics[width=13cm]{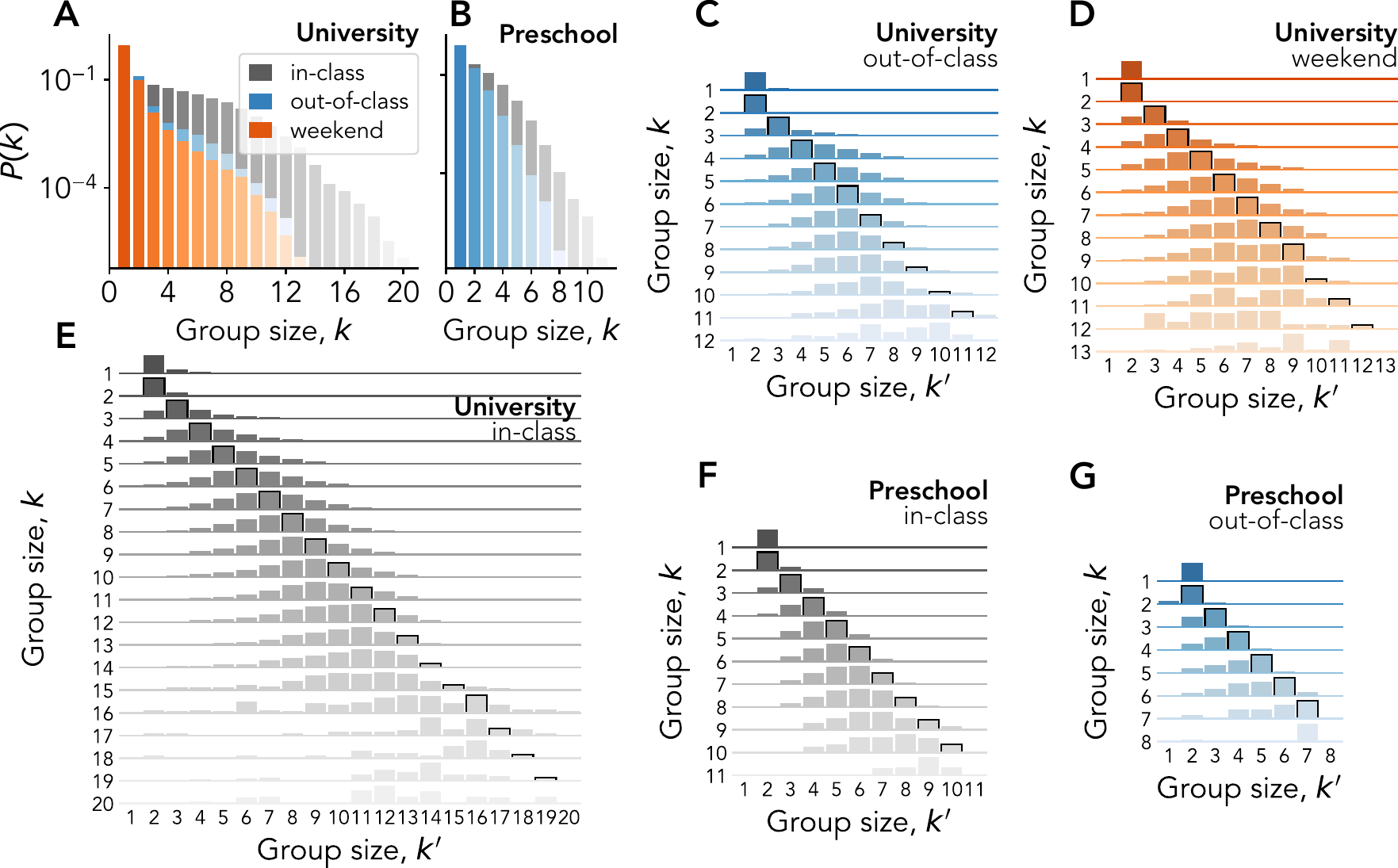}
	\caption[]{
		({\it A-B}) Group size distributions for University ({\it A}) and Preschool ({\it B}) interactions that take place in-class, out-of-class, or during the weekend (see legend).
		({\it C-G}) Node transition matrices for University ({\it C-E}) and Preschool ({\it F-G}), for interactions that take place during in-class ({\it E,F}), out-of-class ({\it C,G}), or weekend ({\it D}) time.
        The elements of each matrix represent the conditional probability that a node that is member of a group of size $k$ at time $t$ is next member of a different group of size $k^{\prime}$ at time $t+1$ ---given that it undergoes a group change between $t$ and $t+1$.
        Probability values are given by the height of each element (normalized by row). Note that the scales on the y-axes ---one for each matrix row--- vary for visualization purposes.
	}
	\label{fig:transition_matrix}
\end{figure*}

A first coarse summary of the complexity of group interactions is unveiled through the heterogeneity of groups sizes, already documented in a variety of studies~\cite{patania2017shape, benson2018simplicial, cencetti2021temporal, iacopini2022group, korbel2023homophily,mancastroppa2023hyper}. We confirm this finding in the data sets considered here in Fig.~\ref{fig:transition_matrix}{\it A,B}, with distributions of instantaneous group sizes having similar shape but spanning varying ranges of values: the interactions measured in the university (panel {\it A}) feature larger group sizes than preschool ones (panel {\it B}), possibly because of the longer range of the Bluetooth signals used in the data collection infrastructure. In addition, we observe a general tendency of gathering in smaller groups in contexts where students or children are free to interact ({\it out-of-class} and {\it weekend}).

\begin{figure}[]
	\centering
	\includegraphics[width=.9\textwidth]{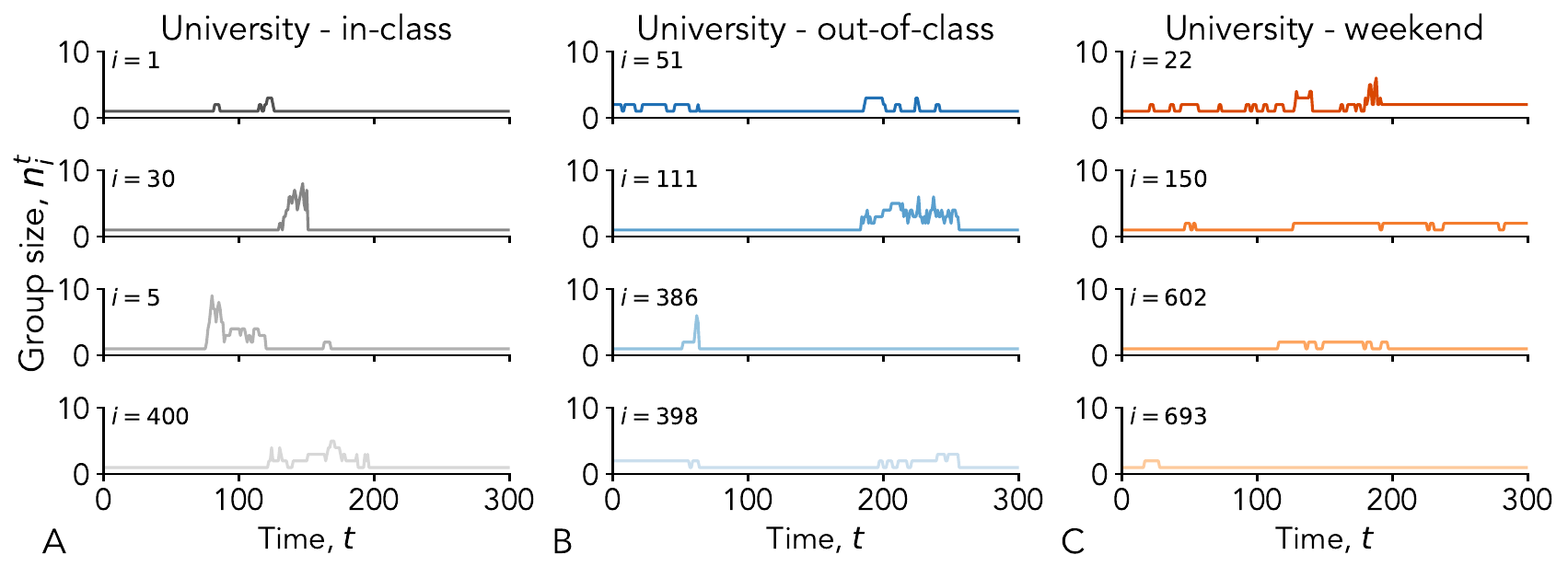}
	\caption[CNS data: groups-size trajectories]{
 Example of timelines for group participation in CNS data divided by context of interaction: in-class ({\it A}), out-of-class ({\it B}), and weekend ({\it C}). Each panel shows a trajectory across different values of group size $n_i^t$ as experienced by a node $i$ at time $t$.
 }
        \label{fig:SI:CNS:trajectories}
\end{figure}

\begin{figure}[]
	\centering
 	\includegraphics[width=.63\textwidth]{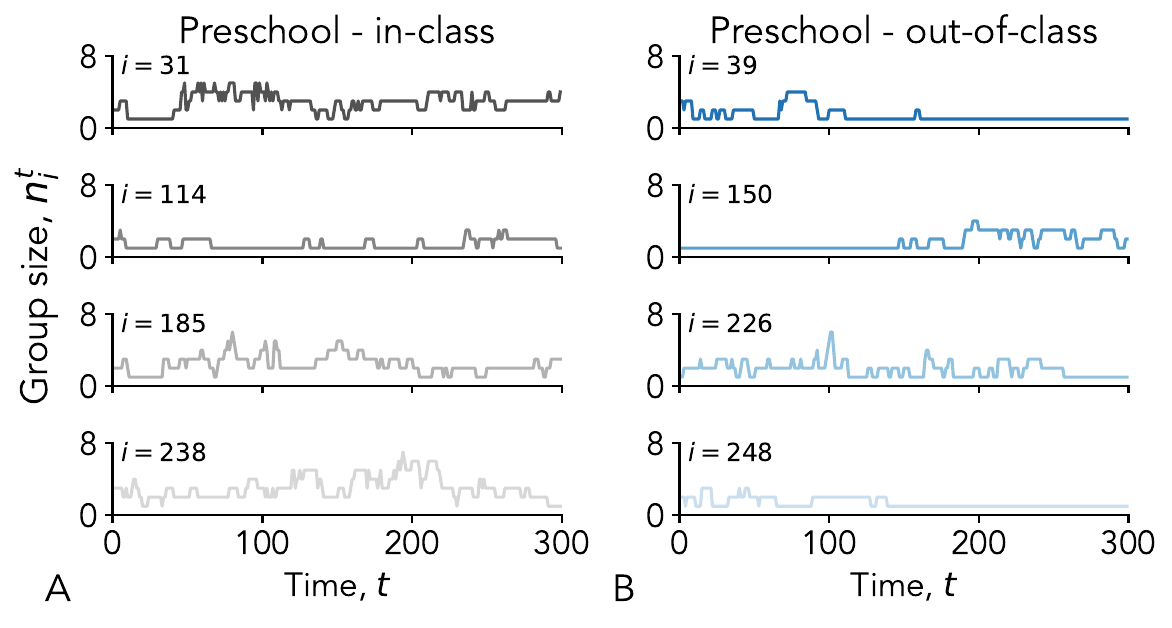}
	\caption[DyLNet data: groups-size trajectories]{
 Example of timelines for group participation in DyLNet data divided by context of interaction: in-class ({\it A}), and weekend ({\it B}). Each panel shows a trajectory across different values of group size $n_i^t$ as experienced by a node $i$ at time $t$.
 }
        \label{fig:SI:DyLNet:trajectories}
\end{figure}

The distribution of sizes is however, by design, an aggregated observable that does not inform us about the dynamics of interactions: a given node might belong at different times to groups of very different sizes, just as a node in a temporal network might have very different numbers of neighbours or centrality values at different times~\cite{braha2006centrality,braha2009time,pedreschi2020dynamic}. 
We thus now investigate how the group membership of individuals evolves across various sizes (some example trajectories can be found in \SI{Fig.~\ref{fig:SI:CNS:trajectories} and Fig.~\ref{fig:SI:DyLNet:trajectories}}).
In this regard, it is important to stress that whenever we refer here to a group change by a node, we interpret it in the most general sense, i.e., it does not necessarily mean that the node is actively changing from a group to another one. In fact, from the point of view of a given individual, a group change can also be due to another person joining or leaving their current group. Under this approach, adopted to avoid having to arbitrarily decide how group ``labels'' propagate whenever there is a change in one of the members, our analysis is purely observational, and agnostic with respect to the intention of individuals.

We build for each context a transition matrix $T\equiv\{T_{kk^{\prime}}\}$ describing these changes as measured in the data: denoting by $n_i^t$ the size of a group to which node $i$ belongs at time $t$, each matrix element represents the conditional probability $P(n_i^{t+1}=k^{\prime}|n_i^t=k)$ of finding a given node $i$ in a group of size $k^{\prime}$ at time $t+1$ given that at time $t$ it belonged to a different group of size $k$ (see {\it Methods}). 
The results, displayed in Fig.~\ref{fig:transition_matrix}{\it C-G}, show strikingly robust patterns across the different contexts, differing only in the cut-off associated with the largest group sizes observed: ({\it i}) at given group size $k$ at $t$, the most probable group size at the next time step is $k'=k$, for small enough $k$ (except for $k=1$ in which case the next size is most often $2$); ({\it ii}) the distribution $P(n_i^{t+1}=k^{\prime}|n_i^t=k)$ extends to values around the diagonal $k'=k$, with both events of individuals undergoing a group change towards a larger or a smaller group but large differences between $k$ and $k'$ are rare; ({\it iii}) as $k$ increases, the distribution shifts to the left of the diagonal, i.e., it becomes increasingly probable that a change of group leads an individual to a group of smaller size. 

\begin{figure}
	\begin{center}
		\includegraphics[width=0.8\textwidth]{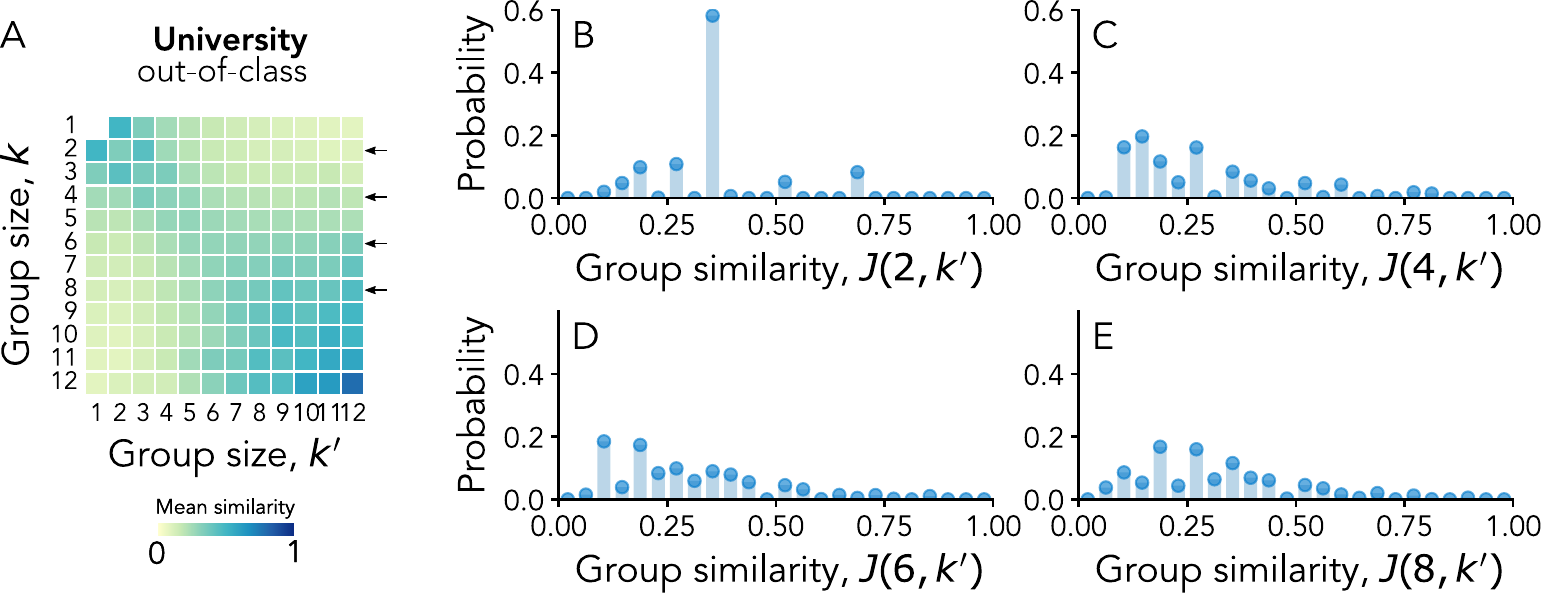}
	\end{center}
	\caption[CNS data: similarity between successive groups]{
		Similarity between successive groups of nodes that undergo a group change for CNS data, out-of-class interactions. ({\it A}) Heatmap of Jaccard similarity between successive groups of a given node, conditioned to the presence of a group change between a group of size $k$ at time $t$ and a group of size $k'$ at time $t+1$. Each value of the matrix represents the similarity averaged across all nodes given two fixed values of $k$ and $k'$. Full distributions (at fixed $k$ but using all possible values of $k'$) for four selected group sizes $k$ are shown in the other panels: $k=2$ ({\it B}), $k=4$ ({\it C}), $k=6$ ({\it D}), and $k=8$ ({\it E}).
	}
	\label{fig:SI:CNS:similarity}
\end{figure}

The approach just described follows the evolution between groups of different sizes from a purely individual standpoint and does not include any information about alter group members beside the considered ego. However, each transition from a group size $k$ to $k^\prime$ could correspond to very different scenarios in terms of group members. To illustrate this point, we  compute the overlap between consecutive groups of an ego, as measured by the Jaccard coefficient between their sets of members. The ego could for instance be at $t$ and $t+1$ in groups formed by totally distinct alters, leading to a low Jaccard coefficient. On the contrary, a group change could also result from another member of the group leaving, in which case the ego would see a strong overlap between the groups at successive times.
Even at fixed $k$ and $k'$, the distributions of Jaccard similarity values between consecutive groups, shown in 
\SI{Fig.~\ref{fig:SI:CNS:similarity}}, confirm in fact that a broad range of intermediate situations are also encountered in terms of change of group members seen by the ego. 
Therefore, to get better insights on the underlying dynamics from a compositional point of view, we now shift the focus of our analysis from individuals to groups.\\

\begin{figure*}[t]
	\begin{center}
		\includegraphics[width=\linewidth]{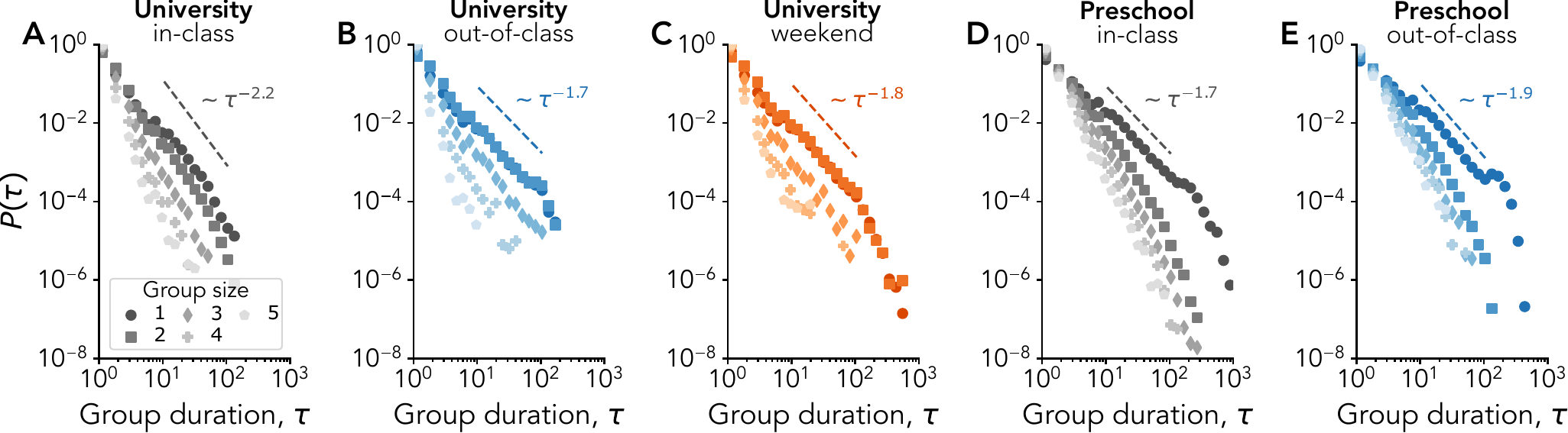}
	\end{center}
	\caption[]{Distributions of group durations $\tau$ for the CNS ({\it A-C}) and the DyLNet ({\it D-E}) data sets in different contexts: in-class ({\it A,D}), out-of-class ({\it B,E}) and weekend ({\it C}). In each panel, different symbols correspond to different group sizes. The distributions for group size $1$ have been fitted using the method in Ref.~\cite{alstott2014powerlaw}, and can be characterized by the depicted exponent values.
	}
	\label{fig:group_duration}
\end{figure*}

For each group, we define its times of birth and death respectively as the first and last appearance of the same set of members in consecutive time steps. The duration of a group is then naturally given by the temporal difference $\tau$ between its time of birth and death, and we investigate in Fig.~\ref{fig:group_duration} how the duration statistics depend on the group size. We find that the  distributions of group duration $P_k(\tau)$ for groups of size $k$ present broad shapes for all $k$, with comparable patterns in shape, exponent values and size dependency across the very different contexts considered.  
In particular, whether students interact during classes, in the other spaces of the university, or elsewhere during the weekend, the distributions of their group interactions depend on the group size in a similar way: the heavy-tail distributions are broader for smaller group sizes, with longer averages and maximum observed durations. Group interactions at preschool show a similar pattern.

\begin{figure*}
	\begin{center}
		\includegraphics[width=\linewidth]{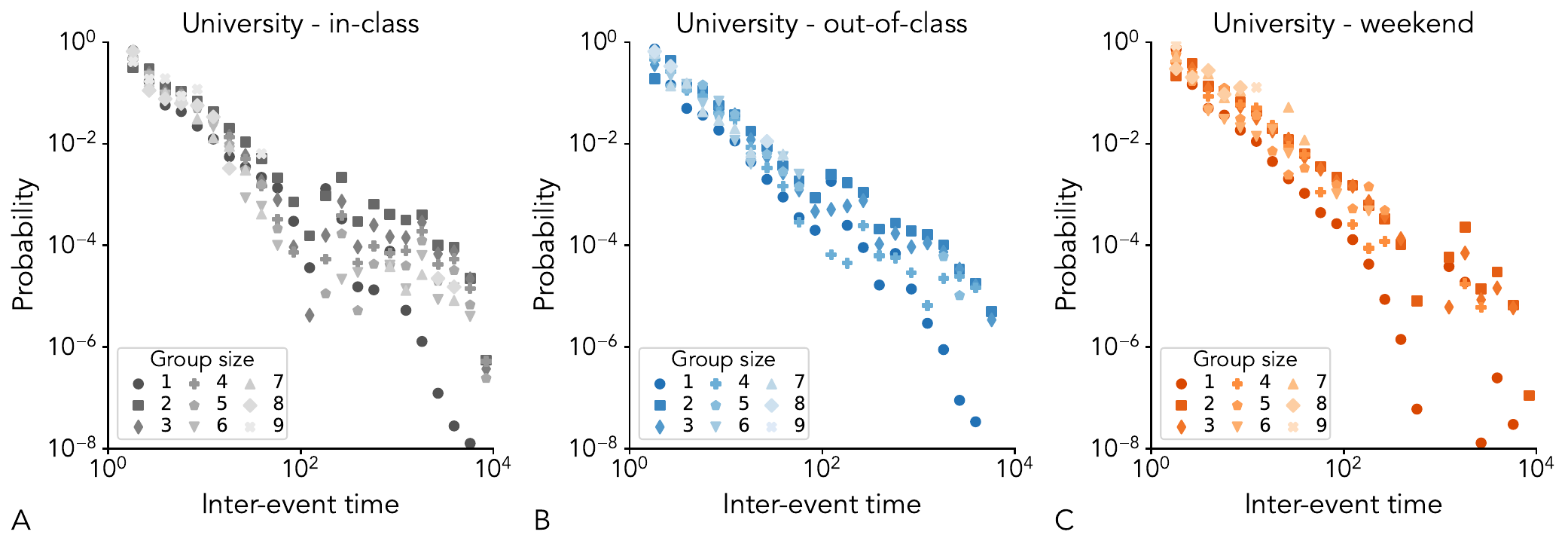}
	\end{center}
	\caption[CNS data: group inter-event times]{
 Distributions of group inter-event times (time between consecutive appearances of the same group) in the CNS data set for different contexts of interaction: in-class ({\it A}), out-of-class ({\it B}), and weekend ({\it C}). In each panel, different symbols correspond to different group sizes.
 }
  	\label{fig:SI:CNS:inter-event_times_dist}
\end{figure*}

\begin{figure*}
	\begin{center}
		\includegraphics[width=0.65\linewidth]{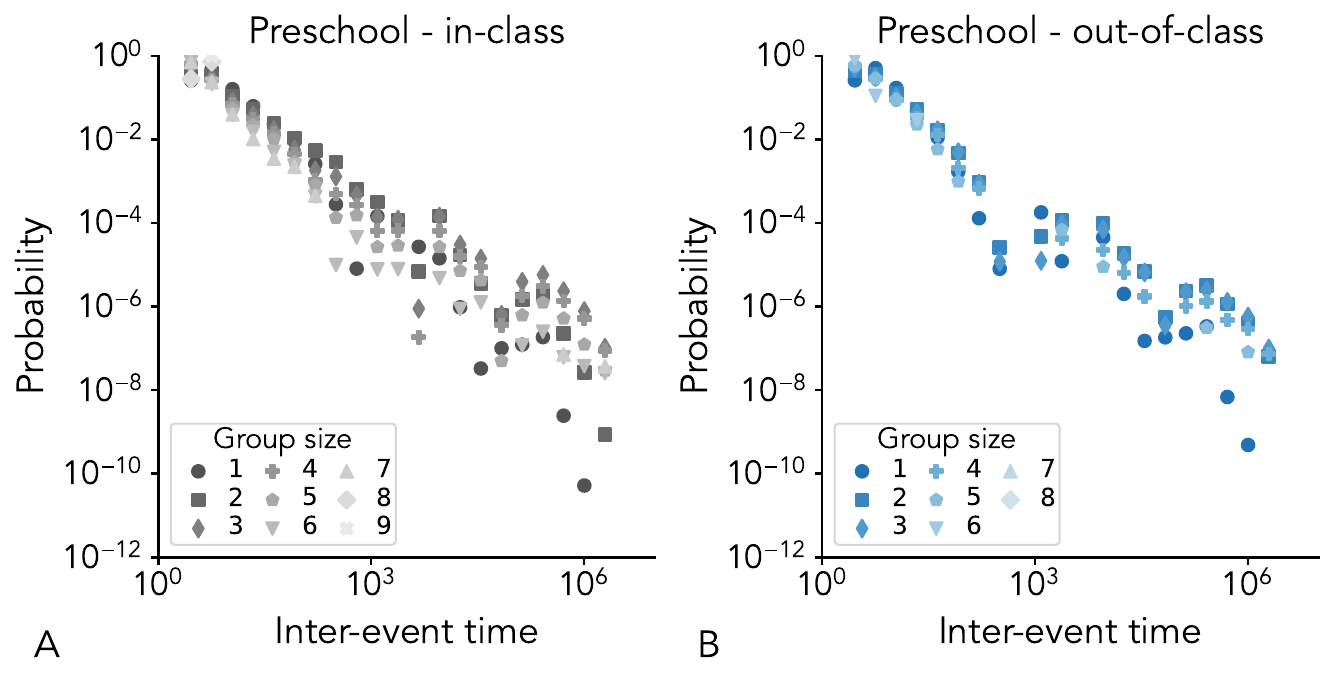}
	\end{center}
	\caption[DyLNet data: group inter-event times]{
 Distributions of group inter-event times (time between consecutive appearances of the same group) in the DyLNet data set for different contexts of interaction: in-class ({\it A}), and out-of-class ({\it B}). In each panel, different symbols correspond to different group sizes.
 }
  	\label{fig:SI:DyLNet:inter-event_times_dist}
\end{figure*}

This phenomenon is closely linked to the one of burstiness, an important feature found in empirical temporal networks~\cite{holme2012temporal}, where periods of node (or link) inactivity are heterogeneously distributed. Moving beyond pairwise interactions, node inactivity corresponds to groups of size 1, i.e., when a node is isolated. Figure~\ref{fig:group_duration} thus confirms the presence of bursty periods of inactivity at the node level. This is also illustrated by examples of node activity through time, as given by the temporal evolution of the size of the group to which a node
belongs, reported for some selected nodes in \SI{Fig.~\ref{fig:SI:CNS:trajectories} and Fig.~\ref{fig:SI:DyLNet:trajectories}}. As expected, nodes display very heterogeneous levels of participation --and active periods featuring medium and large groups are inevitably correlated across different nodes. In addition to the inter-event time distribution for nodes given Fig.~\ref{fig:group_duration}, we also find burstiness across the different contexts of interactions at the level of groups. This is deduced from the inter-event time distributions reported in \SI{Fig.~\ref{fig:SI:CNS:inter-event_times_dist} and Fig.~\ref{fig:SI:DyLNet:inter-event_times_dist}} that are broadly distributed even after disaggregating by group size.
This analysis extends the results described in \cite{zhao2011social,cencetti2021temporal} to very different contexts, showing that the strong robustness of statistical patterns of contacts goes beyond the one of pairwise interactions described in earlier works \cite{cattuto2010dynamics,barrat2014measuring}, and hinting at common robust mechanisms determining contact and group formation, duration, and evolution in different contexts.

\begin{figure*}[t]
	\begin{center}
		\includegraphics[width=0.55\linewidth]{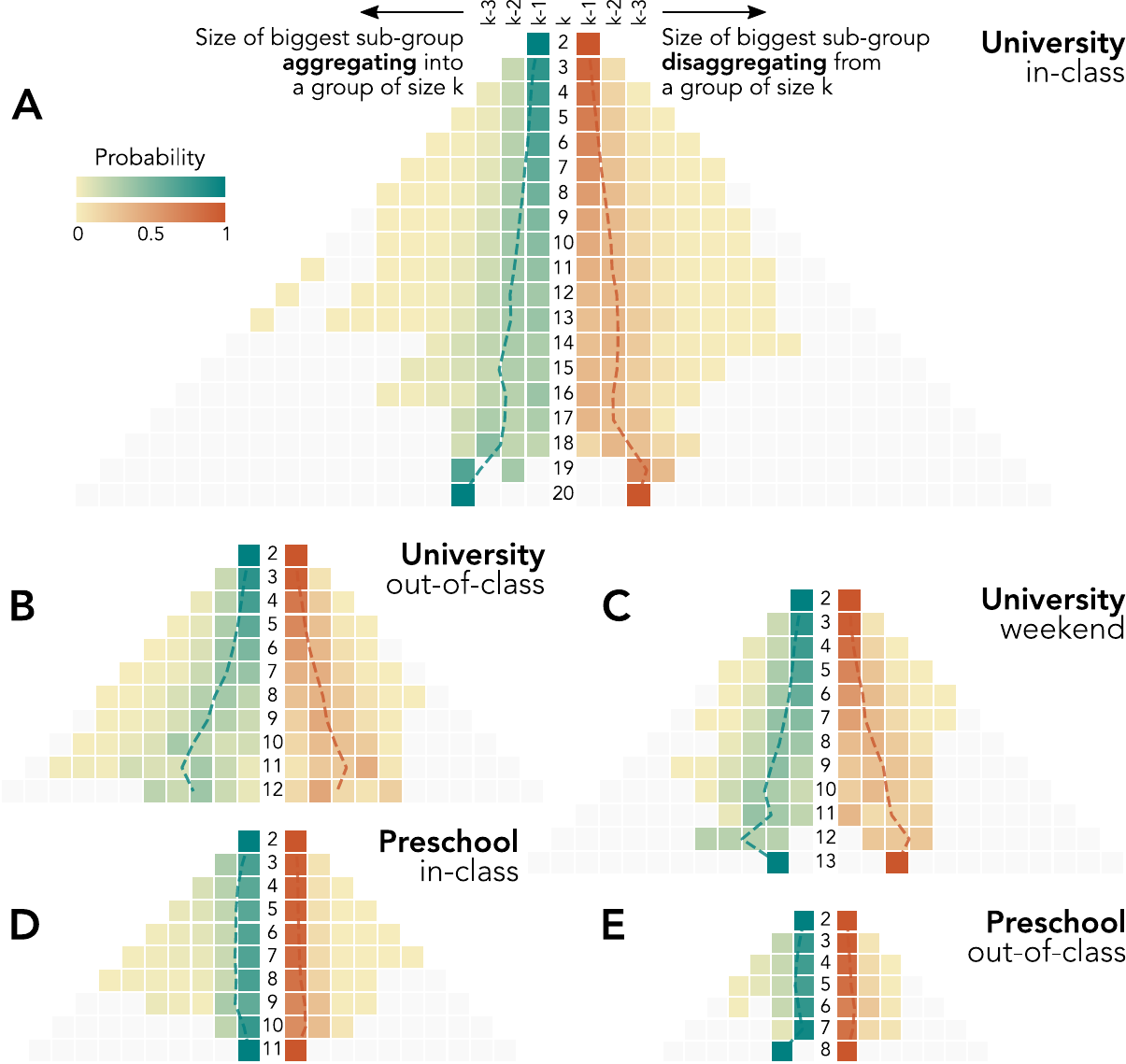}
	\end{center}
	\caption[]{Group dynamics of aggregation and disaggregation for University ({\it A-C}) and Preschool ({\it D-E}) interactions that take place during in-class ({\it A,D}), out-of-class ({\it B,E}), or weekend ({\it C}) time. Each side of the pyramidal heatmaps shows the probability distribution associated to the size for the largest sub-group joining and the largest subgroup leaving a group of size $k$. The central column reports the considered group size $k$, while the probability distributions on its left-hand side and right-hand side respectively corresponds to group aggregation and disaggregation. Dashed lines refer to the distribution average. 
	}
	\label{fig:dis-agg}
\end{figure*}

To go further, we now investigate how groups change: indeed, the node transition matrices introduced above and shown in Fig.~\ref{fig:transition_matrix} give only partial information regarding the actual group dynamics. The individual point of view adopted is useful to understand how individual group membership evolves between different sizes, but the impact of these individual changes on the sizes of the groups needs an independent analysis.
For instance, while Fig.~\ref{fig:group_duration} shows that larger groups tend to have shorter lives, how they break up is still to uncover. Similarly, a group might appear due to the fusion of two pre-existing groups of comparable sizes---like water droplets that merge after overlapping due to surface tension, or from a gradual process of the integration of one individual at a time. To investigate this issue, we follow the members of each group, before the group's birth and after its break-up. Moreover, we pool together the results of groups of the same size, to check whether groups of different sizes undergo different aggregation and disaggregation dynamics. For each group size $k$, we show in Fig.~\ref{fig:dis-agg} the heatmaps (one for each context) of the size distributions of the largest sub-set of group members observed just before the birth or just after the death of a group (see {\it Methods}). 
For small group sizes, both group aggregation and disaggregation tend to happen gradually from or to groups of similar sizes. This is in agreement with previous results for the formation of 3-body interactions~\cite{cencetti2021temporal}. For increasingly larger groups, the picture evolves in a slightly context-dependent way. The general picture is that no merging from (or splitting into) equally sized groups is observed, as could be expected from a purely combinatorial point of view. Medium- and large-sized groups tend to be created from a group of slightly smaller size joined by one or few small ones, and symmetrically lose members in a small chunk, remaining of mid to large size.
This points towards a partially hierarchical dynamical mechanism according to which individuals first engage in small groups and the small to mid and large groups aggregate to form even larger ones. A symmetric process takes place when large groups dismantle ---first into smaller medium-sized subgroups and then loosing members one at the time. 

\subsection*{A data-driven dynamical model for groups' evolution}
Most models describing the temporal evolution of social interactions consider network representations, i.e., are based on mechanisms governing how pairwise interactions are established and successively broken~\cite{perra2012activity, starnini2013modeling, vestergaard2014memory, karsai2014time, nadini2018epidemic,lebail2023modelling}. Here, we describe instead a model that explicitly integrates how individuals form groups of arbitrary sizes \cite{stehle2010dynamical,zhao2011social,petri2018simplicial}:  at each time step an individual can decide to stay in their current group, leave and join a different one, or become isolated. 

The model is inspired by the one put forward in~\cite{stehle2010dynamical,zhao2011social}: we consider $N$ agents that interact in groups over time. 
For simplicity, we assume that each agent $i=1,2,\dots,N$ participates in only one group at a time (as it happens for most of the empirically observed interactions, see \SI{Table~\ref{SI:table:membership}}).
We call $\mathcal{K}^t$ the set of groups present at time $t$. We denote by $\sigma_i^t \in\mathcal{K}^t$ the ---single--- group to which agent $i$ participates at time $t$, of size $n_i^t\equiv|\sigma_i^t|$.
If $n_i^t=1$, the agent $i$ is isolated, or inactive (cf. \SI{Fig.~\ref{fig:SI:CNS:trajectories} and Fig.~\ref{fig:SI:DyLNet:trajectories}}).
We note that many models devoted to describing the evolution of interactions between individuals are based on successive pair interactions (groups of size $2$)~\cite{holme2015modern,perra2012activity,karsai2014time,lebail2023modelling}. 
As time is aggregated, the result of these binary interactions is a social network between a set of nodes representing the individuals~\cite{wasserman1994social}, where a link denotes the fact that two individuals have been in contact (each link can be weighted, e.g., by the cumulated time of interactions). 
In the present case, interactions between agents are described by groups of various sizes. The adequate tool to describe the temporally aggregated picture is then not anymore a network, but a hypergraph between the set of vertices $\mathcal{V}$ representing the agents \cite{hatcher2002algebraic, battiston2020networks}. This hypergraph is formed by hyperedges between these vertices, where a $k$-hyperedge $\sigma\in\mathcal{K}$ is a set of $k$ vertices representing a group interaction of size $k$, which can be weighted by the total time this group interaction has taken place.

The model evolves through iterations where each time step $t$ corresponds to an epoch, during which each one of the $N$ agents is selected in a random order. Whenever an agent $i$ is selected, currently a member of group $\sigma_i^t$, the model then evolves according to two sequential mechanisms. With the first one the agent decides to either stay in the same group, depending on the time spent there and the group size, or alternatively leave it for a different one. If the agent stays, nothing happens. If the agent instead leaves its current group, a second mechanism is triggered, corresponding to the choice of its next group: this choice is based on the acquaintances made until that time. 
We note that in the original pairwise model~\cite{stehle2010dynamical,zhao2011social}, individuals leaving a group became automatically isolated, and isolated individuals could join groups of any size, which implies that the shape of the empirical transition matrix of Fig.\ref{fig:transition_matrix} could not be reproduced. In the next paragraphs we leverage qualitative insights and direct measures from data to 
define these two mechanisms in more details.\\

\begin{figure*}
	\begin{center}
		\includegraphics[width=0.8\linewidth]{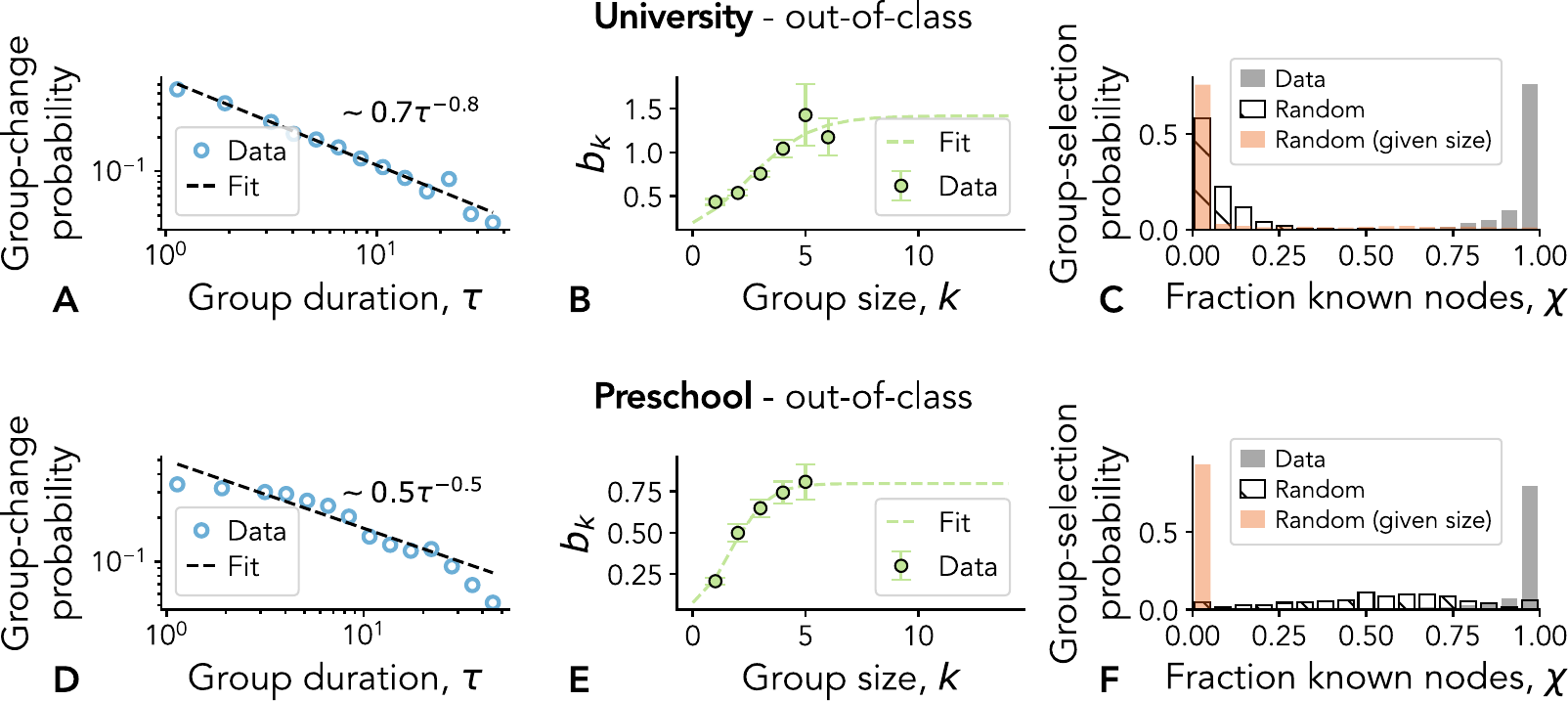}
	\end{center}
	\caption[]{
 Fitting the empirical group-change probability and measuring the signal of social memory for the out-of-class interactions from the CNS ({\it A-C}) and the DyLNet ({\it D-F}) data sets.
 ({\it A,D}) Probability for a node belonging to a group of any size to leave it for a different one after exactly $\tau$ timestamps. Points are binned empirical results, dashed lines represent a power-law fit of the form $b\tau^{\beta}$ ---values reported in each panel. Fitted exponents $\beta$ are then used to estimate the group-size-dependent constants $b_k$, as given in Eq.~\eqref{eq:group_change}, with another power-law fit. In ({\it B,E}) the resulting values for $b_k$ are plotted (points and 95\% CI error bars) together with a logistic fit (dashed lines).
({\it C,F}) Distribution of fraction of nodes ---composing a newly chosen group--- that were previously known to the focal node. The resulting distributions of values, averaged over the different time steps, are plotted in comparison to two null scenarios where the group to join is chosen at random, or at random given the target size.}
	\label{fig:group-change}
\end{figure*}

\begin{figure}
	\centering
	\includegraphics[width=\textwidth]{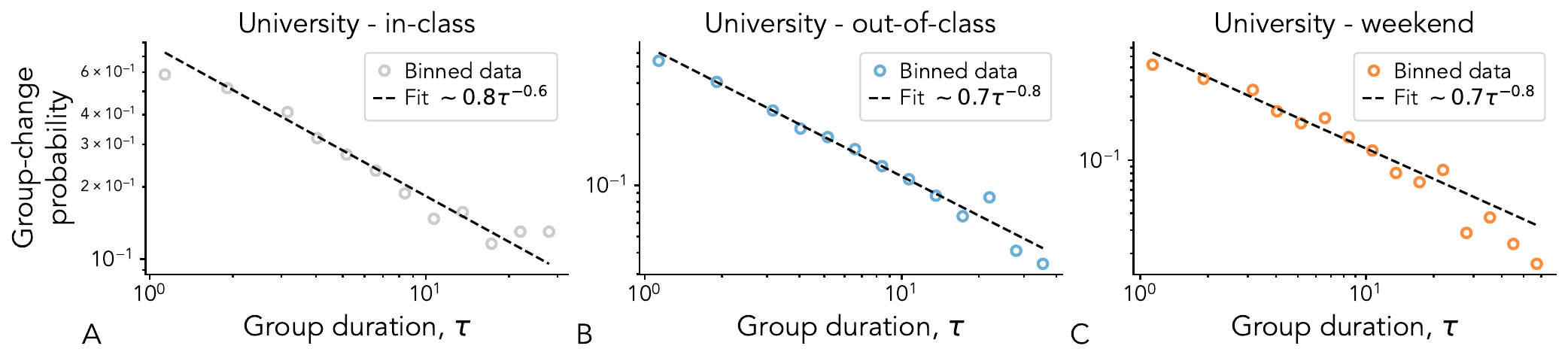}
	\caption[CNS data: group-change probability]{Fitting the group-change probability from the CNS data set by context of interaction: in-class ({\it A}), out-of-class ({\it B}), and weekend ({\it C}). 
		We plot the probability for a node belonging to a group of any size to leave it for a different one after exactly $\tau$ timestamps.
		Points are binned empirical results, dashed lines represent a power-law fit of the form $b\tau^{\beta}$ ---values reported in each panel.
		Compared to Fig.~\ref{fig:SI:CNS:group-change_prob_by_size}, group sizes have been aggregated up to size $k=4$.
	}\label{fig:SI:CNS:group-change_prob}
\end{figure}

\begin{figure}[b]
	\centering
	\includegraphics[width=.7\textwidth]{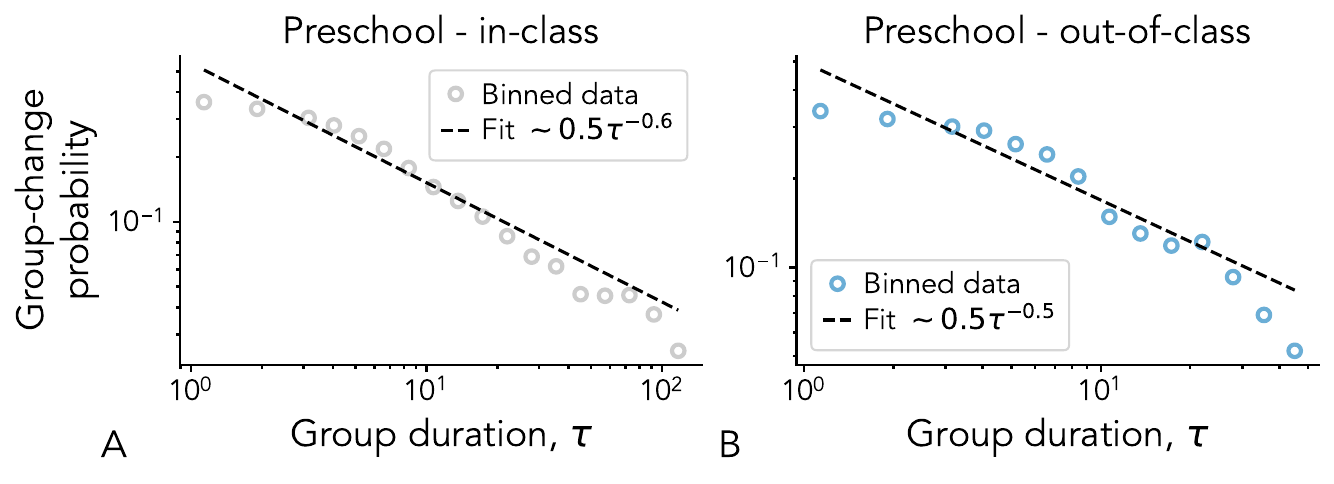}
	\caption[DyLNet data: group-change probability]{Fitting the group-change probability from the DyLNet data set by context of interaction: in-class ({\it A}) and out-of-class ({\it B}). 
		We plot the probability for a node belonging to a group of any size to leave it for a different one after exactly $\tau$ timestamps.
		Points are binned empirical results, dashed lines represent a power-law fit of the form $b\tau^{\beta}$ ---values reported in each panel.
		Compared to Fig.~\ref{fig:SI:DyLNet:group-change_prob_by_size}, group sizes have been aggregated up to size $k=4$.
	}\label{fig:SI:DyLNet:group-change_prob}
\end{figure}

We first take into account that the probability for an agent to have a group change decreases with the time $\tau$ the agent has already spent in that group,  i.e., a ``long-gets-longer'' effect. Evidence for such effect has been found empirically in pairwise interactions \cite{vestergaard2014memory, karsai2014time}. The case of larger group sizes and the potential dependency in the group size have however not been investigated. 
We thus measure the group-change probabilities in our data, and how they depend on the group sizes. Specifically, we compute the probability $P^{\curvearrowright}(k, \tau)$ for a node belonging to a group of size $k$ to have a change of group after $\tau$ timestamps (see {\it Methods}).  
Figure \ref{fig:group-change}{\it A,D} shows the resulting probability distribution as a function of the group duration $\tau$ aggregated over all group sizes $k$ (the distributions shown in the figure correspond to interactions that take place out of class, and results for the other contexts are shown in the \SI{Fig.~\ref{fig:SI:CNS:group-change_prob} and Fig.~\ref{fig:SI:DyLNet:group-change_prob}}). These empirical results suggest that the ``old-gets-older'' mechanism observed in pairwise interactions remains valid for groups: the probability for an agent to change group decreases with the time they have spent in that group. In other words, the longer a group has been established, the smaller the probability that it will break apart. 
Similar trends can be found for the two data sets, both when aggregating over group sizes ---as just shown--- and for the distributions $P^{\curvearrowright}(k, \tau)$ separated by group size, as shown in \SI{Fig.~\ref{fig:SI:CNS:group-change_prob_by_size} and Fig.~\ref{fig:SI:DyLNet:group-change_prob_by_size}} for the CNS and the DyLNet data sets, respectively.

We thus assume in the model that the probability that a node $i$ in a group of size $k$ has a change of group decreases with the time $\tau$ that node $i$ has spent in that group (residence time) as
\begin{equation}\label{eq:group_change}
	p_k(\tau)=\frac{b_k}{\tau^{\beta}/N},
\end{equation}
where $\beta$ is a real-valued exponent that modulates the impact of the residence time, which we obtain by fitting the empirical distributions (dashed lines in Fig. \ref{fig:group-change}{\it A,D}). With probability $1-p_k(\tau)$ the agent $i$ stays in the same group.

\begin{figure}
	\centering
	\includegraphics[width=\textwidth]{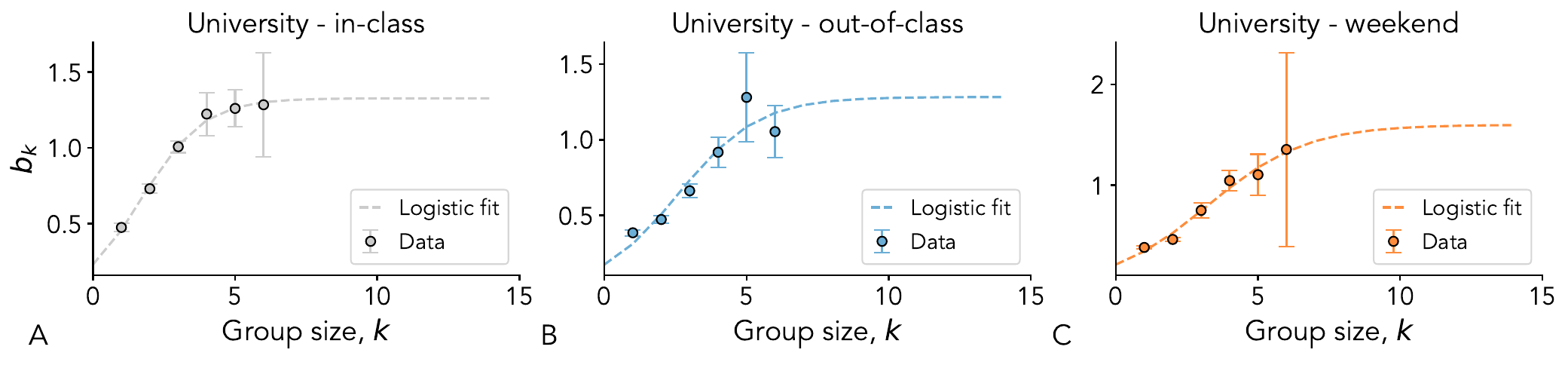}
	\caption[CNS data: estimation of the group-size-dependent constants $b_k$]{Estimation of the group-size-dependent constants $b_k$, as given in Eq.~(1) of the main text, from the CNS data set by context of interaction: in-class ({\it A}), out-of-class ({\it B}), and weekend ({\it C}). For each context, the fitted exponent $\beta$ for the group-change probability reported in Fig.~\ref{fig:SI:CNS:group-change_prob} is used to estimate the values of $b_k$ for different $k$ with power-law fits. Estimated $b_k$ (points and 95\% CI error bars) are plotted together with a logistic fit (dashed lines).
	}\label{fig:SI:CNS:logistic_fit}
\end{figure}

\begin{figure}
	\centering
	\includegraphics[width=.7\textwidth]{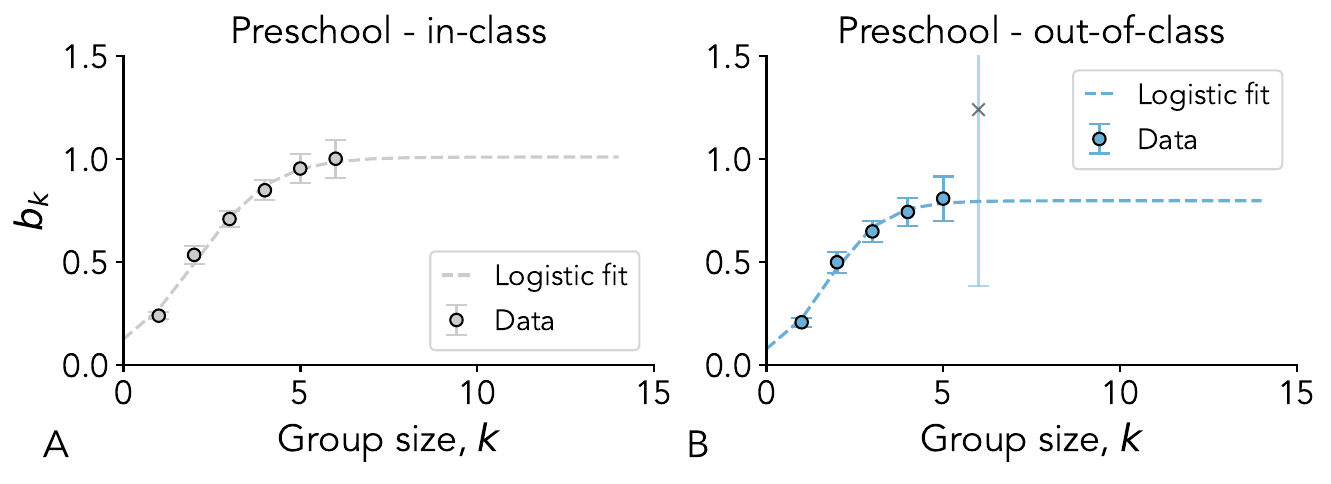}
	\caption[DyLNet data: estimation of the group-size-dependent constants $b_k$]{Estimation of the group-size-dependent constants $b_k$, as given in Eq.~(1) of the main text, from the DyLNet data set by context of interaction: in-class ({\it A}) and out-of-class ({\it B}). For each context, the fitted exponent $\beta$ for the group-change probability reported in Fig.~\ref{fig:SI:DyLNet:group-change_prob} is used to estimate the values of $b_k$ for different $k$ with power-law fits. Estimated $b_k$ (points and 95\% CI error bars) are plotted together with a logistic fit (dashed lines).
	}\label{fig:SI:DyLNet:logistic_fit}
\end{figure}

In Equation \eqref{eq:group_change}, $b_k$ is a constant that depends on the size of the current group of node $i$. It is indeed reasonable to assume that the probability of leaving a group also depends on the size of the group itself, as size is a crucial factor that determines a group's sociological form~\cite{simmel1902number}, and its ability to sustain a single conversation ---leading to the phenomenon known as {\it schisming}~\cite{egbert1997schisming}. To gain insights on the dependency of $b_k$ on the group size $k$, we fit each $P^{\curvearrowright}(k, \tau)$ using a power-law function of the form $b_k \tau^{-\beta}$ (with $\beta$ taken from the fit shown in  Fig. \ref{fig:group-change}{\it A,D}). The resulting values of $b_k$ are reported in Fig.\ref{fig:group-change}{\it B,E}. They show a monotonic increase with $k$, which
we fit to a logistic form
\begin{equation}\label{eq:logistic}
	b_k=\frac{1}{1+e^{-\alpha(k-k_0)}}.
\end{equation}
Similar results across the different contexts and data sets are reported in \SI{Fig.~\ref{fig:SI:CNS:logistic_fit} and Fig.~\ref{fig:SI:DyLNet:logistic_fit}} for the CNS and DyLNet data sets, respectively. In all cases, the logistic fit falls within the confidence intervals of the empirical measures, justifying the choice of a logistic function.

\begin{figure}
	\centering
	\includegraphics[width=\textwidth]{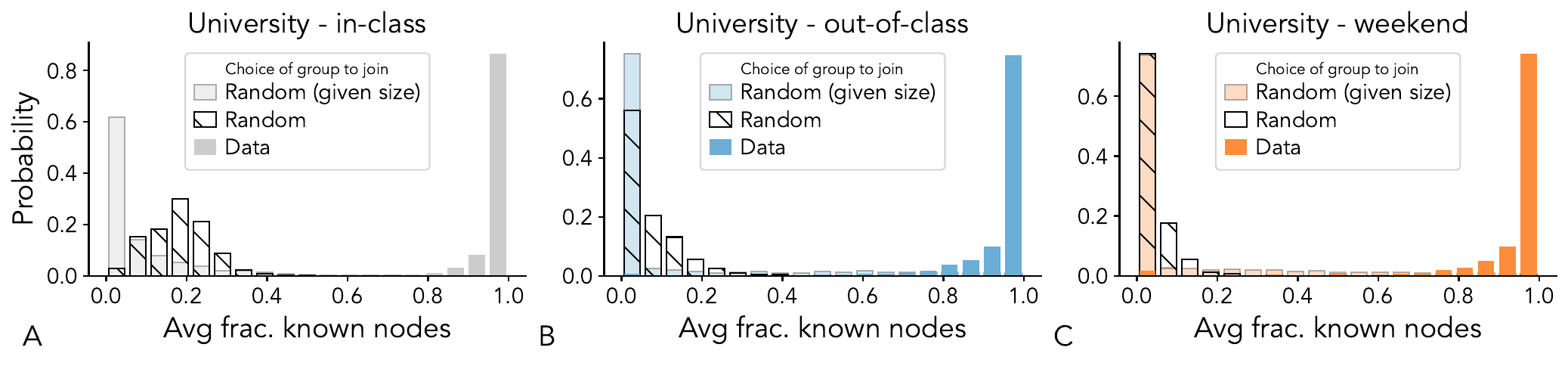}
	\caption[CNS data: measuring social memory]{
 Measuring signals of social memory in group changes for the CNS data set in different contexts of interaction: in-class ({\it A}), out-of-class ({\it B}), and weekend ({\it C}). For each context, whenever we register a group change, we compute the fraction of nodes in the new group that were previously known to the focal node; we also compute the same quantity using a random group of the chosen size available at the considered time, or a random group without any constraints. The resulting distributions of values, averaged over the different time steps, are plotted by comparing the data with the random scenario.
 }
   	\label{fig:SI:CNS:social_memory}
\end{figure}

\begin{figure}
	\centering
	\includegraphics[width=0.7\textwidth]{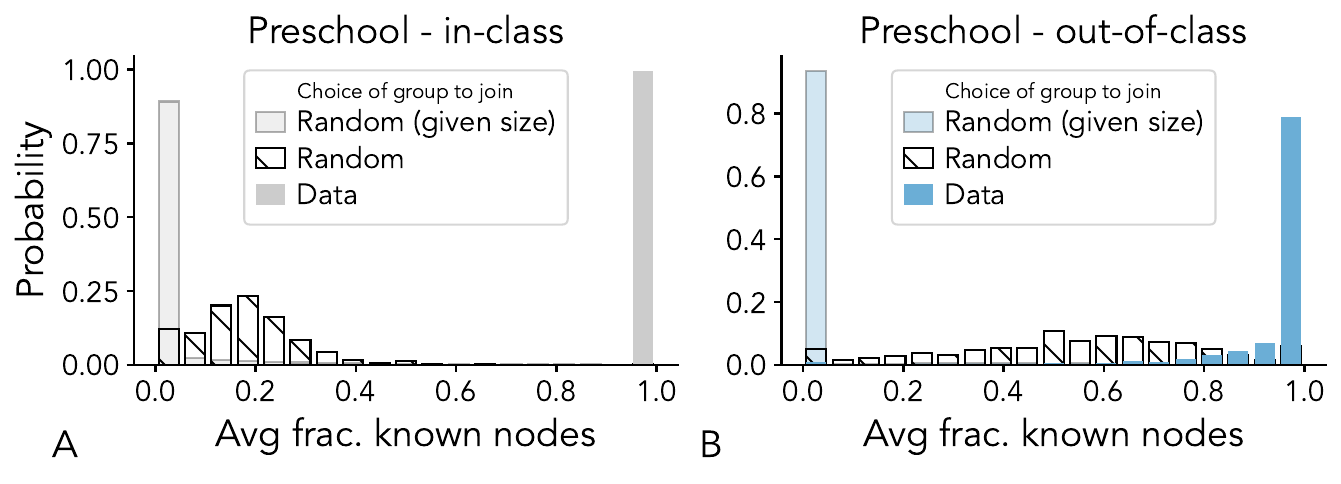}
	\caption[DyLNet data: measuring social memory]{
Measuring signals of social memory in group changes from the DyLNet data set in different contexts of interaction: in-class ({\it A}), and out-of-class ({\it B}). For each context, whenever we register a group change, we compute the fraction of nodes in the new group that were previously known to the focal node; we also compute the same quantity using a random group of the chosen size available at the considered time, or a random group without any constraints. The resulting distributions of values, averaged over the different time steps, are plotted by comparing the data with the random scenario.
 }
        \label{fig:SI:DyLNet:social_memory}
\end{figure}

Let us now focus on the second mechanism of the model, which controls for the selection of the new group after the group change. Previous empirical and modelling investigations for temporal networks support the idea that  individuals have a preference for repeating interactions with people already met, i.e., a mechanism of {\it social memory}~\cite{karsai2014time, vestergaard2014memory,lebail2023modelling}. 
Nevertheless, this explicit signal at the level of groups has never been measured. Hence, we check the presence of social memory in the empirical data by looking, whenever a node $i$ undergoes a group change at any time $t$, at the fraction $\chi_{i,\omega}(t)$ of known nodes in the new group $\omega$ of node $i$. We then take the average over all the group changes and plot the associated distributions for the two considered data sets in Fig.~\ref{fig:group-change}{\it C,F}. We also compare the results with two baseline scenarios in which the group choice is performed uniformly at random, among the available groups at the moment of the change, or at random but restricted to those having the same size of the new group in the empirical data (see {\it Methods}). The results of Fig. \ref{fig:group-change}{\it C,F} clearly show that both data sets display a strong signal of social memory as compared to their random counterparts. The same holds for the other contexts and data sets, as reported in \SI{Fig.~\ref{fig:SI:CNS:social_memory} and Fig.~\ref{fig:SI:DyLNet:social_memory}}.

We thus take into account this mechanism of social memory in the following way in the model. In case of a group change, we denote by $\sigma_i^t$ the group of $i$ at time $t$, and by $\omega\in \Omega_t = \{\mathcal{K}^t\setminus\sigma_i^t\cup\emptyset\}$ the new group after the group change. Note that, by including the empty set $\emptyset$ among the possible target groups, we account for the possibility that $i$ becomes isolated. Specifically, we include in the ensemble $\Omega_t$ of possible groups to join multiple copies of $\emptyset$: this multiplicity, controlled by a parameter $\epsilon$, makes it possible to tune in the model the willingness of agents to become isolated upon leaving a group. Among all possible groups $\omega \in \Omega_t$, $i$ selects the one to join via the second behavioural mechanism, which involves the memory of previous interactions. Namely, the probability to join $\omega$ is proportional to the fraction $\chi_{i,\omega}(t)$ of agents in $\omega$ that at time $t$ have already interacted with $i$ in the past. Let us know define a slightly different quantity $\chi^{\dag}_{i,\omega}(t)$ as
\begin{equation}\label{eq:new_group_density}
	\chi^{\dag}_{i,\omega}(t)=
	\frac{1+\Big[\omega\cap\bigcup\limits_{t'=1}^t\sigma_i^{t'}\Big]}{1+|\omega|}.
\end{equation}
Note that, differently from the simple fraction of known agents, node $i$ itself is included in the computation of $\chi^{\dag}_{i,\omega}(t)$ in order to have a non-zero probability for $i$ to join either an empty group or a group of previously unmet individuals. In other words, $\chi^{\dag}_{i,\omega}(t)$ is the density of agents in the group which are known to $i$ right after joining. Altogether, the probability for an agent $i$ belonging to group $\sigma_i^t$ at time $t$ to be found in a different group $\omega$ at time $t+1$ is given by
\begin{equation}\label{eq:new_group_probability}
	\text{Prob}(\sigma_i^{t+1}=\omega|\{\sigma_i^t\}_t) = \frac{\chi^{\dag}_{i,\omega}(t)}{\sum_{\omega^{\prime}}\chi^{\dag}_{i,{\omega^{\prime}}}(t)}.
\end{equation}

\subsection*{The model reproduces the higher-order dynamical features}

We now explore the ability of the model defined above to capture the key empirical features we have uncovered in the dynamics of group interactions. As empirical results are robust across data sets and contexts, we consider as an example the University interactions taking place during out-of-class time. 
We thus run the model initialised with $N=700$ agents for $T=2000$ time steps, using different parameter values for $\epsilon$, $\alpha$, and $k_0$ ---while $\beta$ is set to $0.8$ as measured in Fig.~\ref{fig:group-change}{\it A}. Each realisation of the model generates a sequence of temporally-ordered hypergraphs that we can analyse as per the empirical data, obtaining in particular group size distributions and group size transition matrices. As described in {\it Methods}, we can thus jointly fit the model on these two observables.

\begin{figure*}[t]
	\begin{center}
		\includegraphics[width=0.55\linewidth]{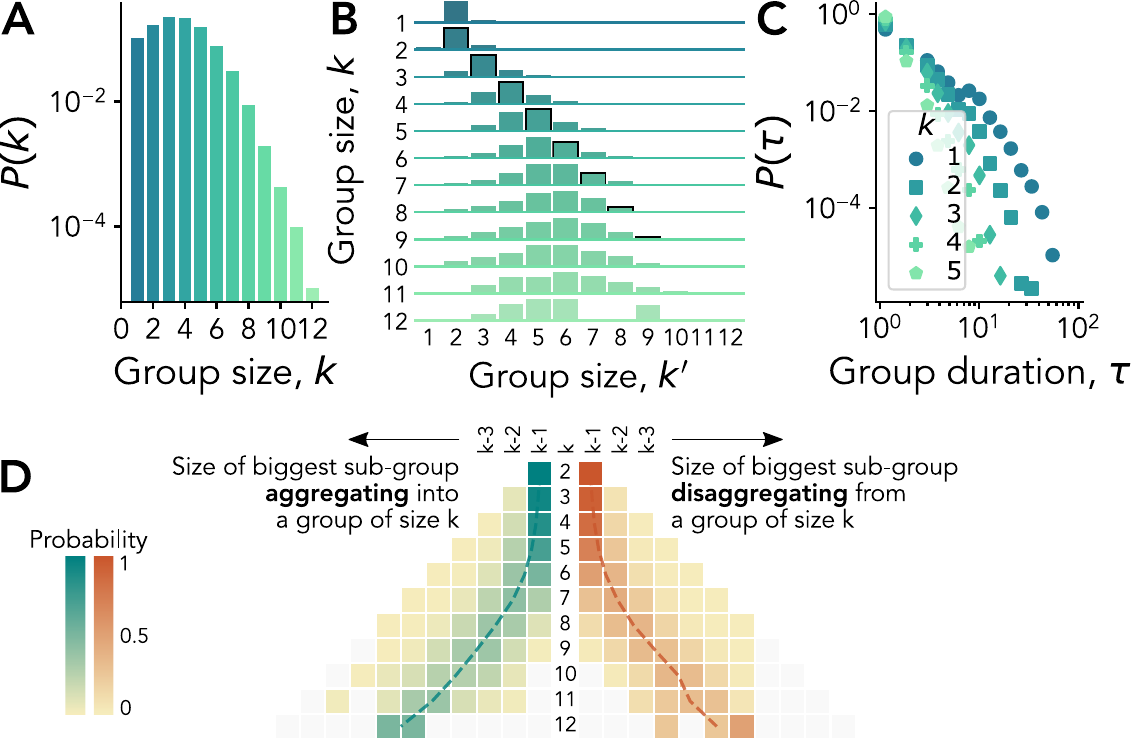}
	\end{center}
	\caption[]{Simulated distribution of group size ({\it A}), node transition matrix ({\it B}), group duration for different group sizes $k$ ({\it C}), and pyramidal heatmap associated to the aggregation and disaggregation dynamics ({\it D}) generated by the proposed temporal hypergraph model.}
	\label{fig:model}
\end{figure*}

We show the results of the best performing model in Fig.~\ref{fig:model}. All obtained results are in line with the empirical data analysed. The group-size distribution (Fig.~\ref{fig:model}{\it A}) spans a range of values comparable with the empirical observation in Fig.~\ref{fig:transition_matrix}{\it A}. The group size transition matrix for the dynamics of group changes from the node point of view, shown in Fig.~\ref{fig:model}{\it B}, has similar symmetric patterns for small sizes 
and a biased transition towards smaller sizes due to the cut-off effect for larger groups, as in Fig.~\ref{fig:transition_matrix}{\it C}. 
It is important to note that other group properties, albeit not taken into account for the exploration of the model parameters, are also reproduced. Indeed, the group duration distributions (Fig.~\ref{fig:model}{\it C}) display broad tails, with a similar group size dependency as in the empirical observations of Fig.~\ref{fig:group_duration}{\it B}. 
More importantly, even complex dynamical characters such as the group disaggregation and aggregation probability distributions, displayed in Figure~\ref{fig:model}{\it D}, resembles the empirical findings (Fig.~\ref{fig:dis-agg}{\it B}), showing the excellent capacity of the proposed model to account for and reproduce the complex phenomenology of the dynamics of group interactions.

\section*{Discussion}
\begin{figure}
	\centering
	\includegraphics[width=\textwidth]{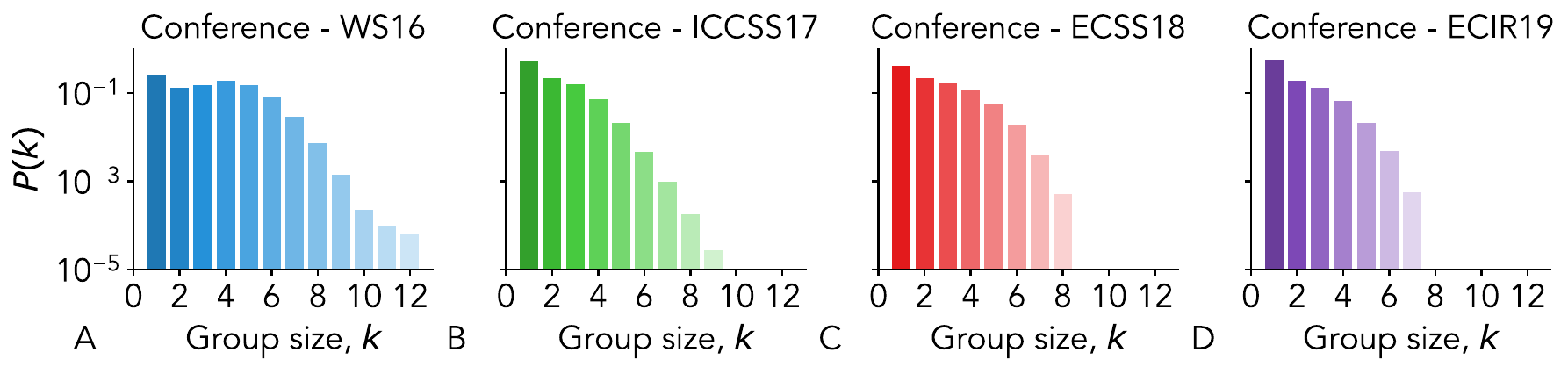}
	\caption[GESIS data: group size distributions]{
Group size distributions from the GESIS data set for interactions taking place during different events: WS16 ({\it A}), ICCSS17 ({\it B}), ECSS18 ({\it C}), and ECIR19 ({\it D}).
 }
        \label{fig:SI:Confs:group-size_dist}
\end{figure}

\begin{figure}
	\centering
	\includegraphics[width=\textwidth]{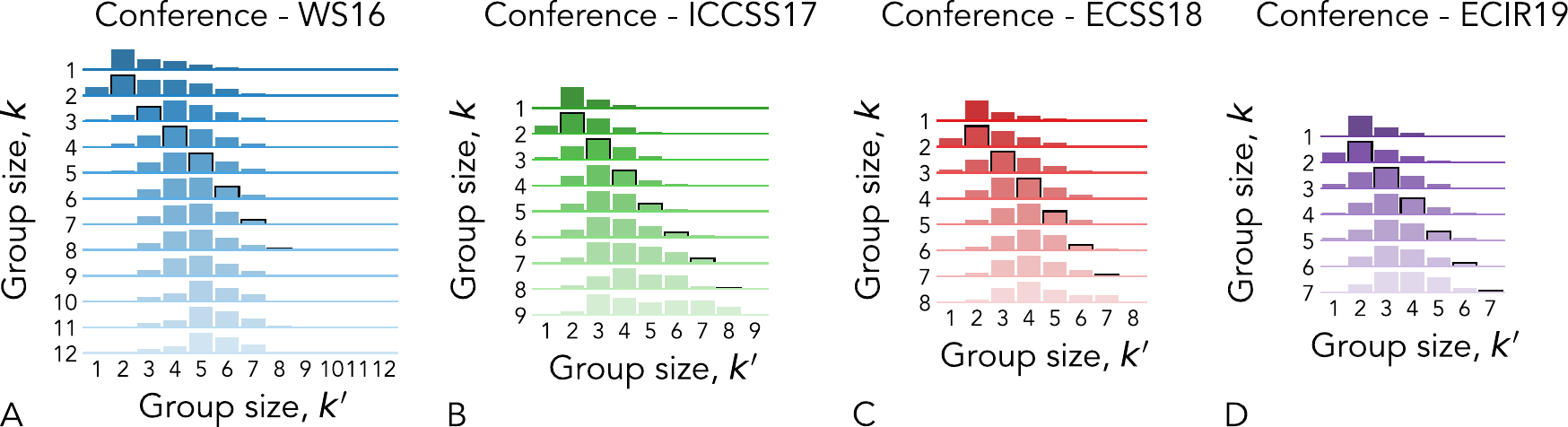}
	\caption[GESIS data: node transition matrices]{
Node transition matrices from the GESIS data set for interactions taking place during different events: WS16 ({\it A}), ICCSS17 ({\it B}), ECSS18 ({\it C}), and ECIR19 ({\it D}). The elements of each matrix represent the conditional probability that a node that is member of a group of size $k$ at time $t$ is next member of a different group of size $k^{\prime}$ at time $t+1$ ---given that it undergoes a group change. Probability values are given by the height of each element (normalized by row). Note that the scales on the y-axes ---one for each matrix row--- vary for visualization purposes.
 }
         \label{fig:SI:Confs:transition_matrices}
\end{figure}

\begin{figure*}
	\begin{center}
		\includegraphics[width=\linewidth]{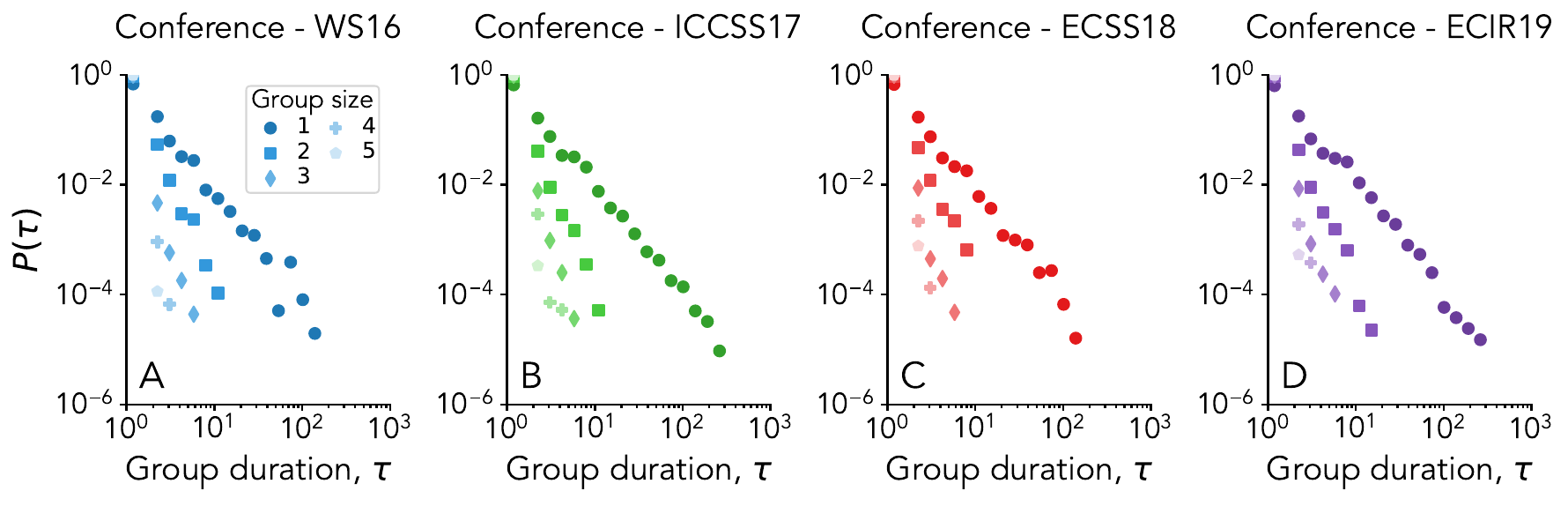}
	\end{center}
	\caption[GESIS data: distributions of group durations]{
Distributions of group durations $\tau$ for the GESIS data set in different contexts: WS16 ({\it A}), ICCSS17 ({\it B}), ECSS18 ({\it C}), and ECIR19 ({\it D}). In each panel, different symbols correspond to different group sizes.
 }
         \label{fig:SI:Confs:group-duration_dist}
\end{figure*}

\begin{figure*}
	\begin{center}
		\includegraphics[width=0.6\linewidth]{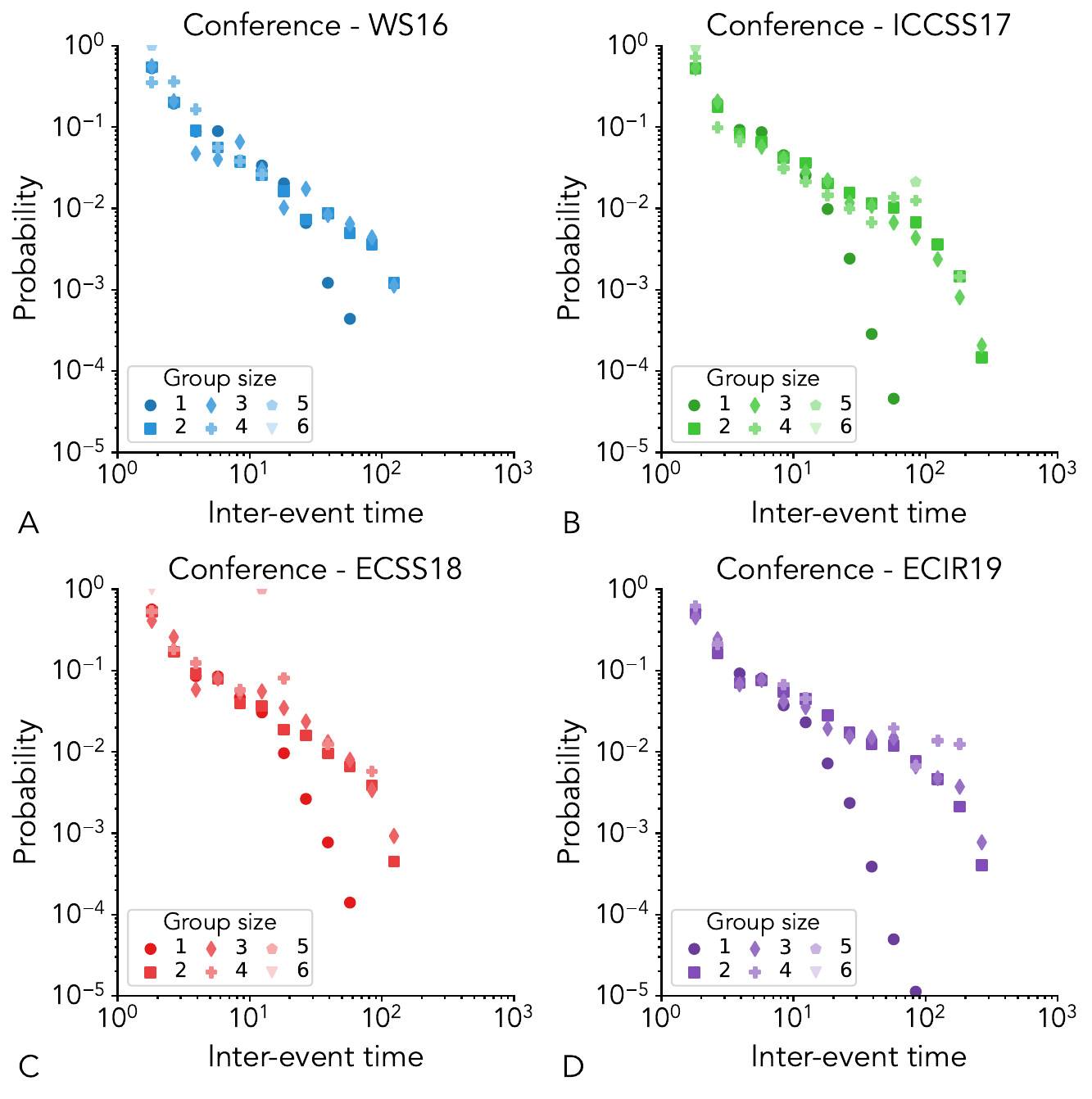}
	\end{center}
	\caption[GESIS data: distributions of inter-event times]{
Distributions of group inter-event times (time between consecutive appearances of the same group) in the GESIS data set for different contexts of interaction: WS16 ({\it A}), ICCSS17 ({\it B}), ECSS18 ({\it C}), and ECIR19 ({\it D}). In each panel, different symbols correspond to different group sizes.
 }
         \label{fig:SI:Confs:inter-event_time_dist}
\end{figure*}

\begin{figure*}
	\begin{center}
		\includegraphics[width=0.5\linewidth]{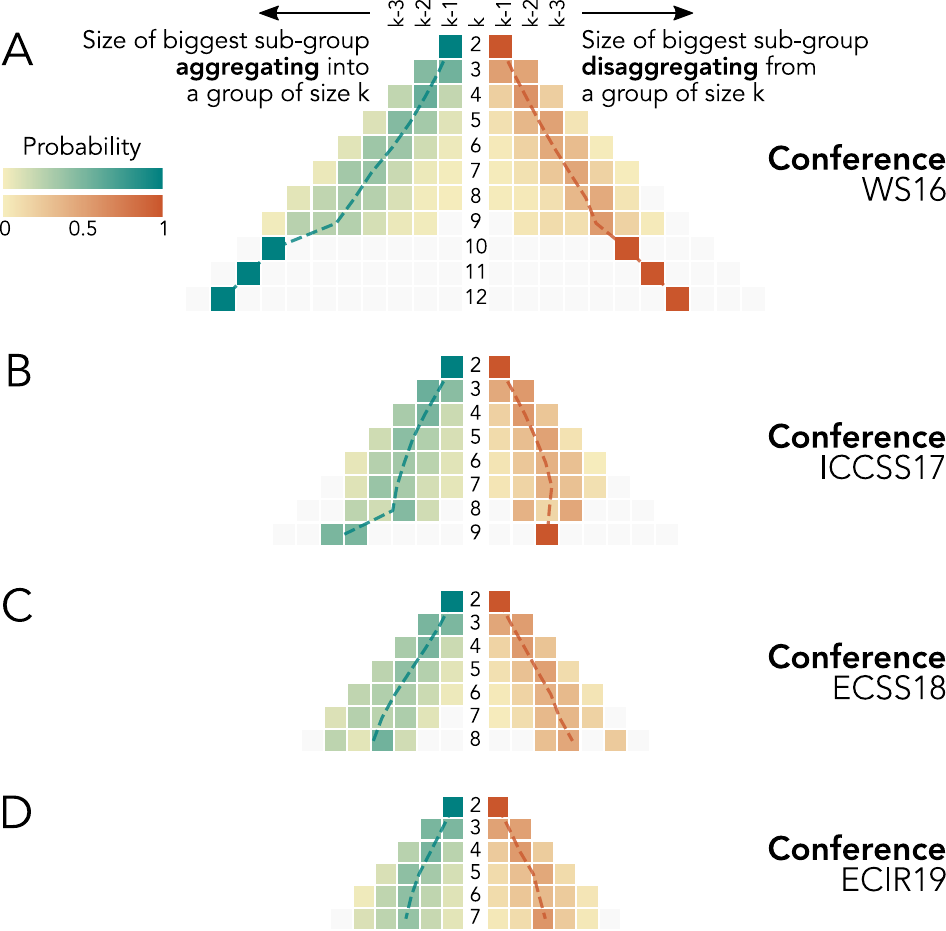}
	\end{center}
	\caption[GESIS data: group dynamics of aggregation and disaggregation]{
Group dynamics of aggregation and disaggregation from the GESIS data set for interactions taking place during different events: WS16 ({\it A}), ICCSS17 ({\it B}), ECSS18 ({\it C}), and ECIR19 ({\it D}). Each side of the pyramidal heatmaps shows the probability distribution associated to the size for the largest sub-group joining and the largest subgroup leaving a group of size $k$. The central column reports the considered group size $k$, while the probability distributions on its left-hand side and right-hand side respectively corresponds to group aggregation and disaggregation. Dashed lines refer to the distribution average.
}
         \label{fig:SI:Confs:group_agg-dis}
\end{figure*}

\begin{figure}
	\centering
	\includegraphics[width=0.7\textwidth]{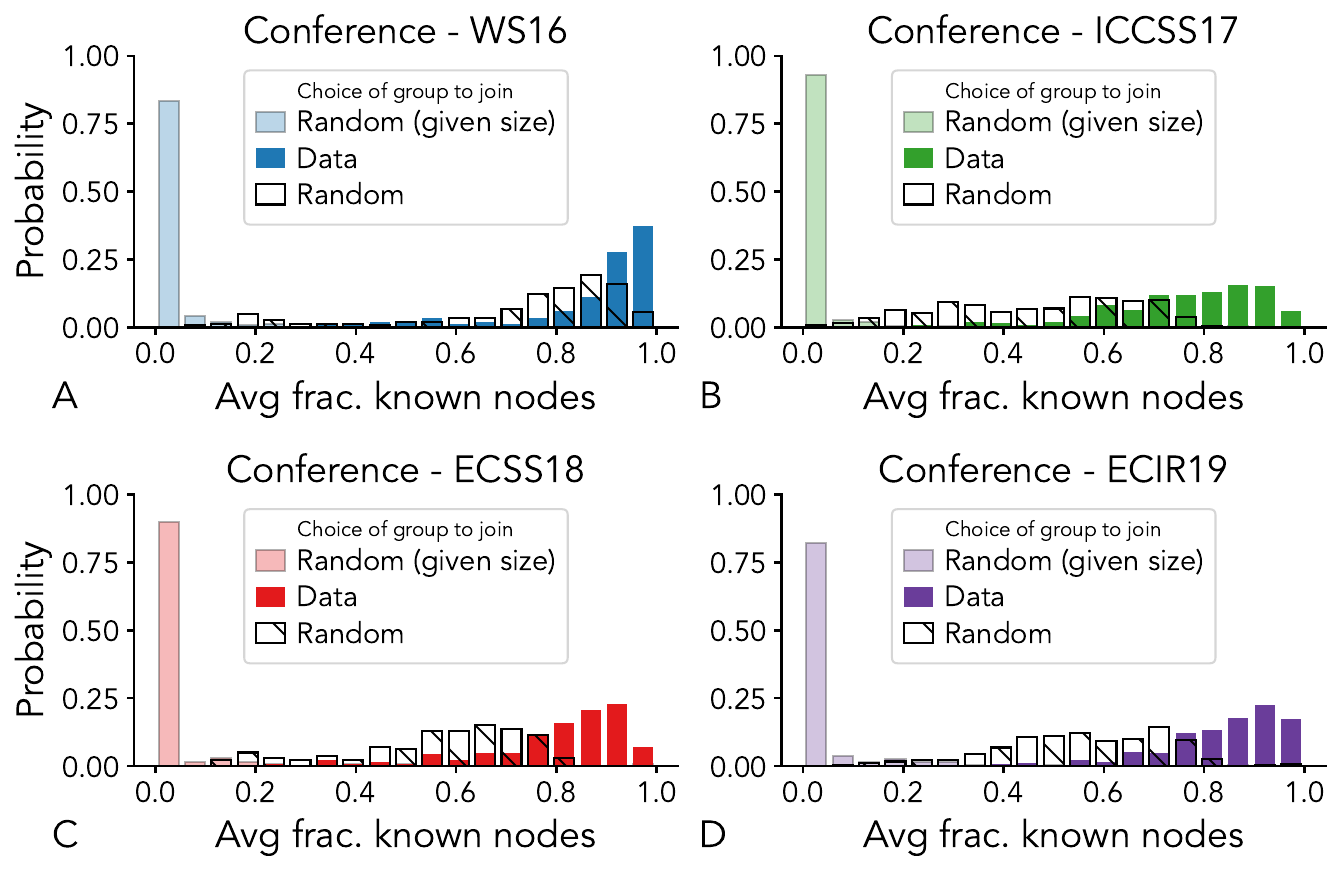}
	\caption[GESIS data: measuring social memory]{
Measuring signals of social memory in group changes for GESIS data collected in different context of interaction: WS16 ({\it A}), ICCSS17 ({\it B}), ECSS18 ({\it C}), and ECIR19 ({\it D}). For each context, whenever we register a group change, we compute the fraction of nodes in the new group that were previously known to the focal node; we also compute the same quantity using a random group of the chosen size available at the considered time, or a random group without any constraints. The resulting distributions of values, averaged over the different time steps, are plotted by comparing the data with the random scenario.
 }
 	\label{fig:SI:Confs:social_memory}
\end{figure}

We have here analysed human interactions under the lens of group dynamics in two data sets, collected respectively among preschool children and university freshmen students. Despite the inherently different nature of their interactions due to age, contexts, and setting constraints such as class schedules, and despite the differences in data collection techniques, we have uncovered strikingly similar group dynamics both at the individual and group level. In particular, we have observed similar group size and duration distributions, and more importantly, consistent dynamical patterns of individual group transitions and group formation and dissolution phenomena in the two settings and at times corresponding to different activity types.
Strikingly, we show in the Supplementary Information that these signatures can also be found in data collected in other contexts, such as social interactions that took place during four different scientific conferences organised by GESIS~\cite{genois2023combining} (see \SI{Note \ref{SI:sec:Confs} and Figs.~\ref{fig:SI:Confs:group-size_dist}, \ref{fig:SI:Confs:transition_matrices}, \ref{fig:SI:Confs:group-duration_dist}, \ref{fig:SI:Confs:inter-event_time_dist}, \ref{fig:SI:Confs:group_agg-dis}, \ref{fig:SI:Confs:social_memory}}). 
The individuals whose interactions are reported in those data are more heterogeneous than in the university and preschool scenarios analysed here: conference participants cover indeed a broad range of ages, with interactions involving different levels of seniority. Even the context can be considered as a mixture of the in-class and out-of-class settings, as conference schedules usually provide more freedom than classes ---with participants often not attending all sessions. Despite all these differences, very similar patterns of group statistics and group changes are observed. It would of course be interesting to extend these results even further by considering still other contexts of human interactions to confirm the generality of such group dynamics and patterns of group size change among humans. 

\begin{figure}
	\centering
	\includegraphics[width=0.35\textwidth]{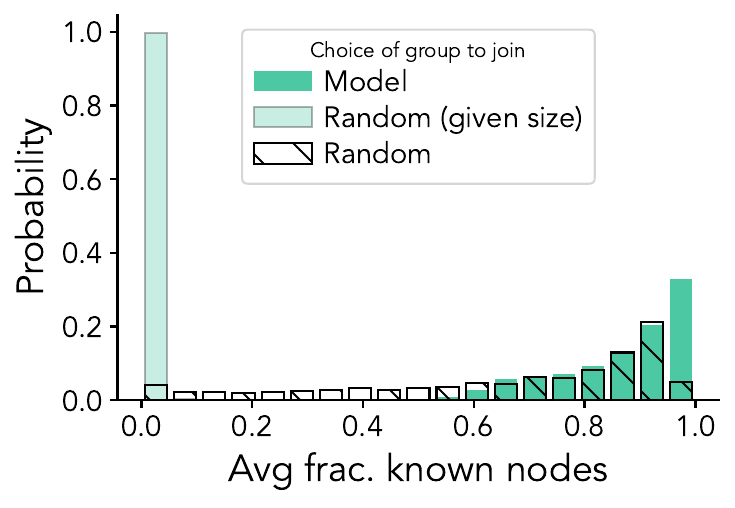}
	\caption[Synthetic model: measuring social memory]{
Measuring signals of social memory in group changes in a simulation of the proposed temporal hypergraph model. For each time step of the simulation, whenever we register a group change, we compute the fraction of nodes in the new group that were previously known to the focal node; we also compute the same quantity using a random group of the chosen size available at the considered time, or a random group without any constraints. The resulting distributions of values, averaged over the different time steps, are plotted by comparing the data with the random scenario.
 }
        \label{fig:SI:model:social_memory}
\end{figure}

Our analysis and results contribute to the obtention of a more complete representation of social dynamics than the ones limited to pairwise interactions. We have accordingly proposed a synthetic model describing how nodes representing individuals form groups and navigate between groups of different sizes. The model includes mechanisms of short-term memory (``long gets longer'') and long-term social memory (higher probability to join a group including individuals already encountered, see \SI{Fig.~\ref{fig:SI:model:social_memory}}), and is able to reproduce the non-trivial dynamics of group changes, both from a node-centric point of view and from the point of view of group formation and break-up. 
Note that, both when discussing the robustness of the patterns obtained in different empirical contexts, and when validating the model, we have remained at a phenomenological level, for several reasons. First, there is no clearly recognized quantitative measure of distance or similarity between two temporal networks or temporal hypergraphs. Second, the data sets we study have different sizes and maximal group sizes, so that the matrices we look at have different sizes, and the distributions have different cut-offs. Third, we prefer to avoid claims about specific shapes of functions (especially for distributions with broad shapes) or of universality, as such claims are notoriously difficult to establish or disprove. Finally, even if we were to define an ad hoc quantitative measure of difference combining the various observables in an arbitrary manner, we would not have a reference value to compare it to.

Thanks to its realistic group dynamics, our model could be used to generate synthetic substrates for studying the impact of higher-order temporal interactions on dynamical processes. Indeed, while the impact of higher-order interactions on various dynamical processes has been well assessed \cite{iacopini2019simplicial,battiston2021physics}, studies on structures undergoing a realistic temporal evolution are scarce \cite{Chowdhary_2021, shang2023non}.
The interplay with the dynamics of groups might prove relevant for a wide variety of processes of interest in many contexts, such as the dynamics of adoption~\cite{barrat2022social} and opinion formation~\cite{neuhauser2022consensus}, but also in synchronisation~\cite{skardal2022explosive, millan2022geometry}, cooperation \cite{Traulsen2006Evolution, sheng2024strategy, civilini2024explosive} and other evolutionary dynamical processes~\cite{perc2013evolutionary}.
Overall, our results call for the development of more modelling approaches that explicitly take both the temporal and the many-body nature of social interactions into account, both to understand the mechanisms from which the complex group dynamics emerges, and to investigate the consequences of such group gatherings in collective dynamics. 
For instance, the model could also be extended to other forms of memory, as explored in pairwise interactions~\cite{lebail2023modelling}. In addition, the model we implemented relies on a set of minimalistic mechanisms that shape the behaviour of the nodes. Future work may further enrich these rules ---for the choice of group change and selection--- by integrating homophily-driven decisions and mechanisms of opinion dynamics that would co-evolve together with the social structure~\cite{schweitzer2022social}.
Notably, all these approaches should not be limited to humans, as non-human animals have also shown to be sensible to higher-order social effects~\cite{katz2011inferring, rosenthal2015revealing, iacopini2024not}, and, at the pairwise interaction level, complex features similar to the ones of human interactions have been observed~\cite{Gelardi2020MeasuringSN}.
Further studies would however be required in order to integrate behavioural response to non-pairwise interactions with additional environmental~\cite{flierl1999individuals}, cultural~\cite{conradt2000activity}, and ecological factors ---like splitting for resource competition, or grouping as a defensive strategy against predators~\cite{wittemyer2005socioecology}. Indeed, there are cases in which the drivers of animal grouping can have genetic roots~\cite{archie2006ties}.
Alternatively, one might try to devise a microfounded principle that could explain the observed temporal evolution of group sizes in term of balancing costs and benefits ---akin to evolutionary models used for collective action problems~\cite{gavrilets2015collective}. Along this line, a game-theoretic interpretation has been given for the different group sizes of static hypergraphs constructed from scientific collaborations~\cite{alvarez2021evolutionary}.

Our results inevitably rely on the given definition of group interaction, as constructed from pairwise data, which represents a proxy for real-world group encounters that also depend on the temporal resolution of the data. Given the current lack of a longitudinal data collection effort specifically designed to track group interactions,  several research directions could be explored. For example, it would be interesting to check the robustness of the empirical results with respect to other definitions of groups or hyperedges from data obtained by measuring pairwise interactions, such as Bayesian inference approach to distinguish hyperedges from combinations of lower-order interactions \cite{young2021hypergraph}, or extraction of statistically significant hyperedges \cite{musciotto2021detecting}. Even the hardcore definition of group that we used could be challenged, using instead less stringent conditions~\cite{alvarez2023inference} together with the possibility of having nodes that display multi-membership.

To conclude, our study contributes towards a better understanding of human behaviour in terms of the formation and disaggregation of groups, and of the navigation between social groups. We expect that the analysis presented can support researchers working at the intersection of social and behavioural sciences, while the proposed model can directly be used to inform more realistic simulations of social contagion, norm emergence, and spreading phenomena~\cite{vespignani2012modelling} in interacting populations.

\section*{Methods}
\subsection*{Data Description and pre-processing}

\subsubsection*{Copenhagen Network Study}

We use data collected via the Copenhagen Network Study (CNS) \cite{sapiezynski2019interaction} that represents a temporally-resolved proximity data collected through the Bluetooth signal of cellular phones carried by 706 freshmen students at the {\it Technical University of Denmark}. 
The publicly available data corresponds to the data recorded during four weeks of a semester, and describes proximity of students with a temporal resolution of 5 minutes. The raw data (already pre-processed in Ref.~\cite{sapiezynski2019interaction}) contains 5,474,289 records. Each entry contains a timestamp, the ID of one user (ego), the ID of another user (alter), and the associated Received Signal Strength Indication (RSSI) measured in dBm. The data is already processed to neglect the directionality of each interaction (which device is scanning). Empty scans (no other device found) are reported with a 0 RSSI, which corresponds to isolated nodes.

We split the data records into three main periods according to the hour of the day and the day of the week. Even though the released data~\cite{sapiezynski2019interaction} do not contain precise information on time and date, these can be easily inferred by cross-checking activity patterns in the temporal sequences with the official timetables for Bachelor studies at DTU~\cite{DTU_corse_base}. The resulting contexts are:
\begin{itemize}
	\item {\it Workweek (in-class)}: Monday to Friday, 8 a.m. to 5 p.m.
	\item {\it Workweek (out-of-class)}: Monday to Friday, 12 a.m. (midnight) to 8 a.m. and 5 p.m. to 11.59 p.m.
	\item {\it Weekend}: Saturday and Sunday.
\end{itemize}

\begin{figure}
	\bigskip
	\begin{center}
		\includegraphics[width=0.8\linewidth]{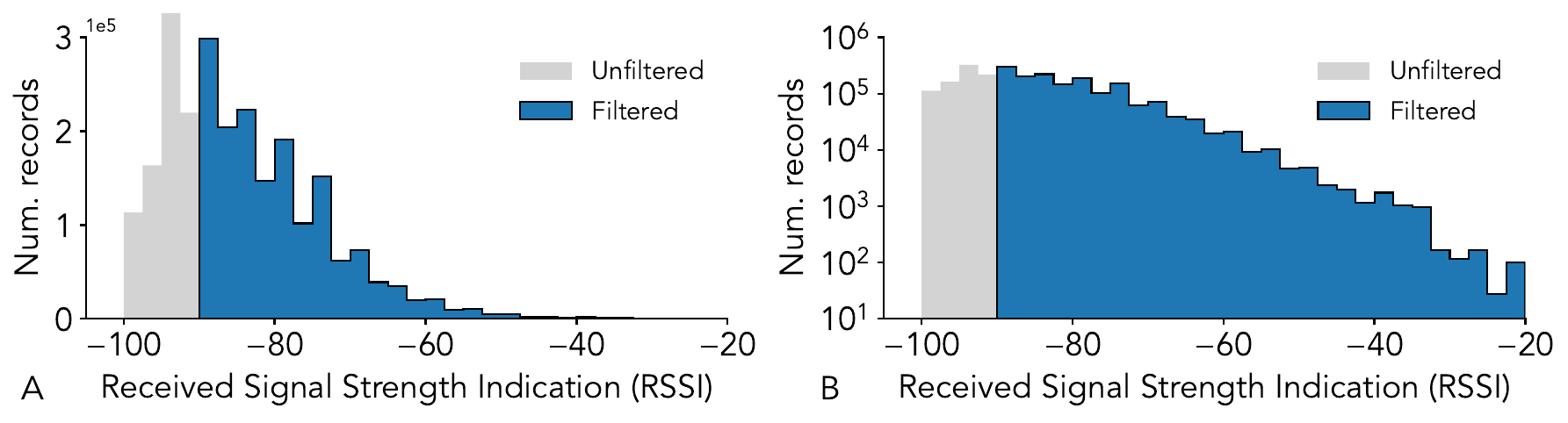}
	\end{center}
	\caption[CNS data: filtering by RSSI]{Filtering bluetooth signals from the CNS data set. ({\it A}) Number of recorded interactions as a function of the associated Received Signal Strength Indication (RSSI) [dBm]. The y-axis in panel ({\it B}) is in logarithmic scale. The volume of interactions retained after the filtering is displayed in blue.
	}\label{fig:SI:CNS:RSSI_filtering}
\end{figure}

We further clean the data in the following way. First, we remove external users by deleting all records in which the device of a participant scanned a device that did not take part in the experiment (resulting in 4,646,415 records). 
We then retain only records with an RSSI higher than -90dBm [see \SI{Fig.~\ref{fig:SI:CNS:RSSI_filtering}}]. This is slightly less restrictive than the threshold of -80dBm used in \cite{sekara2014strength}, which was used to select interactions occurring within a radius of 2 meters (a typical distance for social interactions among close acquaintances \cite{hall1990hidden}). 
After doing this, we have 3,824,052 records divided into 1,603,916 pairwise interactions and 2,220,136 empty scans. We treat the latter as isolated nodes.

We then perform three pre-processing steps as in Ref.~\cite{sekara2016fundamental}. First, 
in order to smooth the pairwise interactions, we  look for all the gaps composed by pairwise interactions that are present at times $t-1$ and $t+1$ but not at time $t$. We fill the resulting 163,349 gaps by using the mean RSSI of the adjacent timestamps (eventually replacing, if present, a record of an empty scan from one of the two interacting nodes). 

Second, we filter out spurious interactions by removing all the 130,935 pairwise signals that are present solely at time $t$ but not at times $t-1$ and $t+1$, leaving us with 3,855,139 records. This is also in line with the procedure performed in Ref.~\cite{sekara2016fundamental}, which is based on the convention developed by the {\it Rochester Interaction Record}~\cite{reis1991studying}, according to which an encounter needs to last 10 minutes or longer to be classified as meaningful. 

\begin{figure}
	\bigskip
	\begin{center}
		\includegraphics[width=0.4\linewidth]{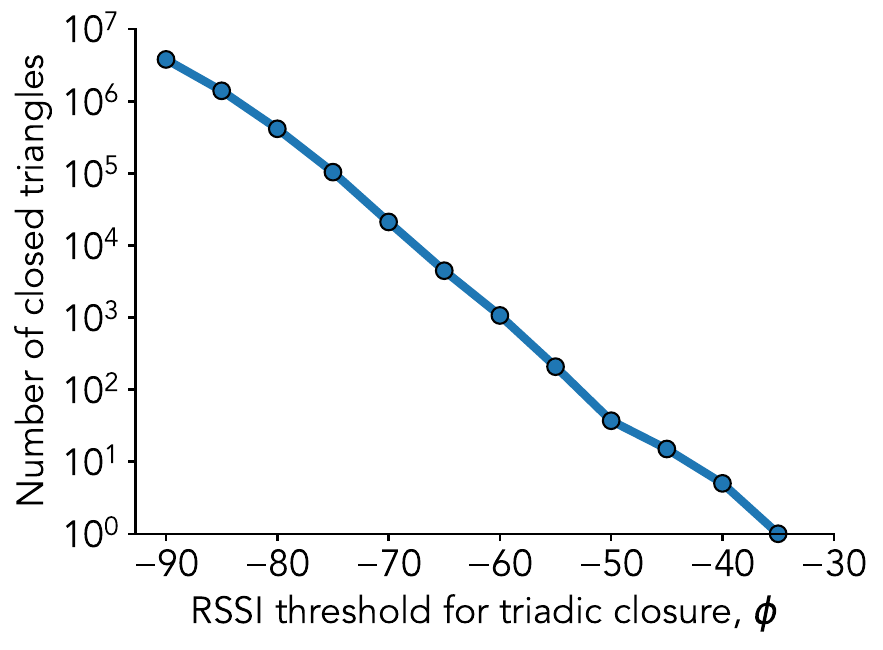}
	\end{center}
	\caption[CNS data: triadic closure vs RSSI]{Pre-processing of the CNS dataset: triadic closure. Number of links added via triadic closure as a function of their approximated Received Signal Strength Indication (RSSI).
	}\label{fig:SI:CNS:tri-clo_RSSI}
\end{figure}

\begin{figure}
	\bigskip
	\begin{center}
		\includegraphics[width=0.5\linewidth]{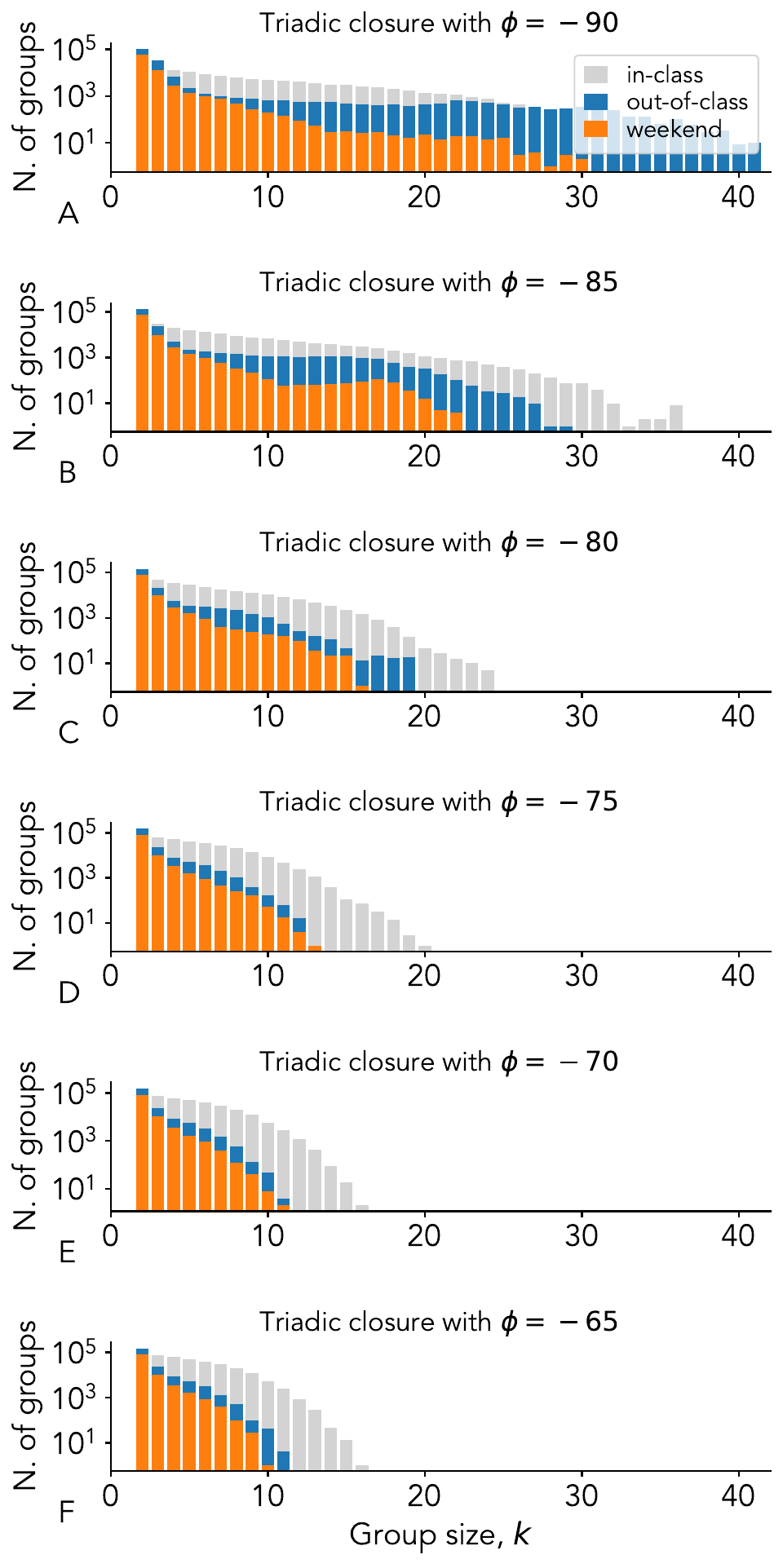}
	\end{center}
	\caption[CNS data: impact of thresholding for the triadic closure]{Pre-processing of the CNS dataset: triadic closure with threshold.
		Impact of thresholding the newly added link ---due to the triadic closure--- on the group size distributions for different contexts of interaction. Each panel shows the number of groups, divided by context [in-class (gray), out-of-class (blue), weekend (orange)], as a function of the group size $k$. Different panels (rows) correspond to different value of the threshold $\phi$ used to filter out the links added with the triadic closure. Links featuring a Received Signal Strength Indication (RSSI) [dBm] lower than $\phi$ are thus removed.
	}\label{fig:SI:CNS:tri-clo_thresholds}
\end{figure}

\begin{figure}
	\bigskip
	\begin{center}
		\includegraphics[width=0.85\linewidth]{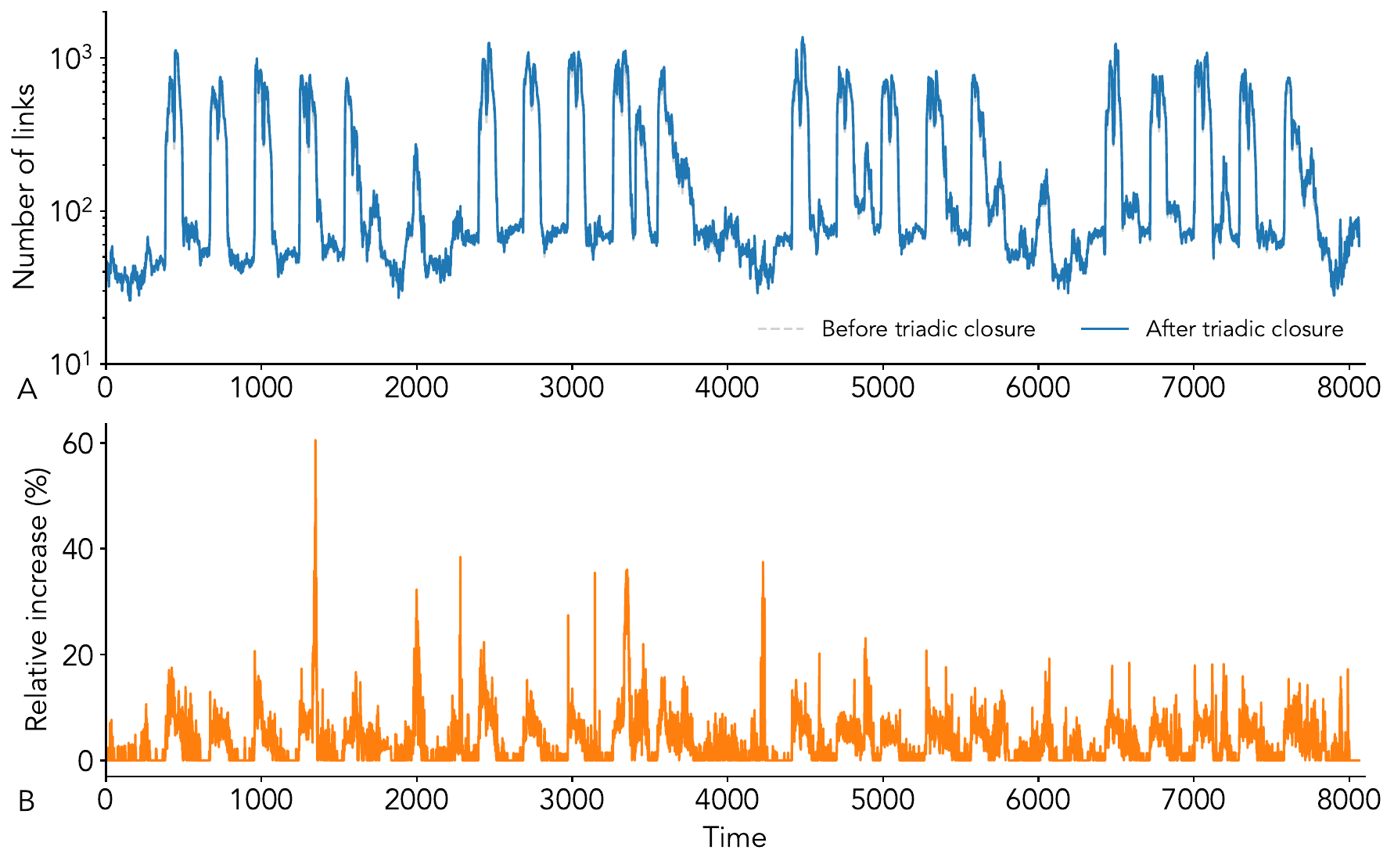}
	\end{center}
	\caption[CNS data: impact of triadic closure on number of interactions]{Preprocessing of the CNS dataset: impact of the triadic closure on the number of raw pairwise interactions in time ({\it A}), and relative change ({\it B}).
	}\label{fig:SI:CNS:tri-clo_num_interactions}
\end{figure}

\begin{figure}
	\bigskip
	\begin{center}
		\includegraphics[width=0.85\linewidth]{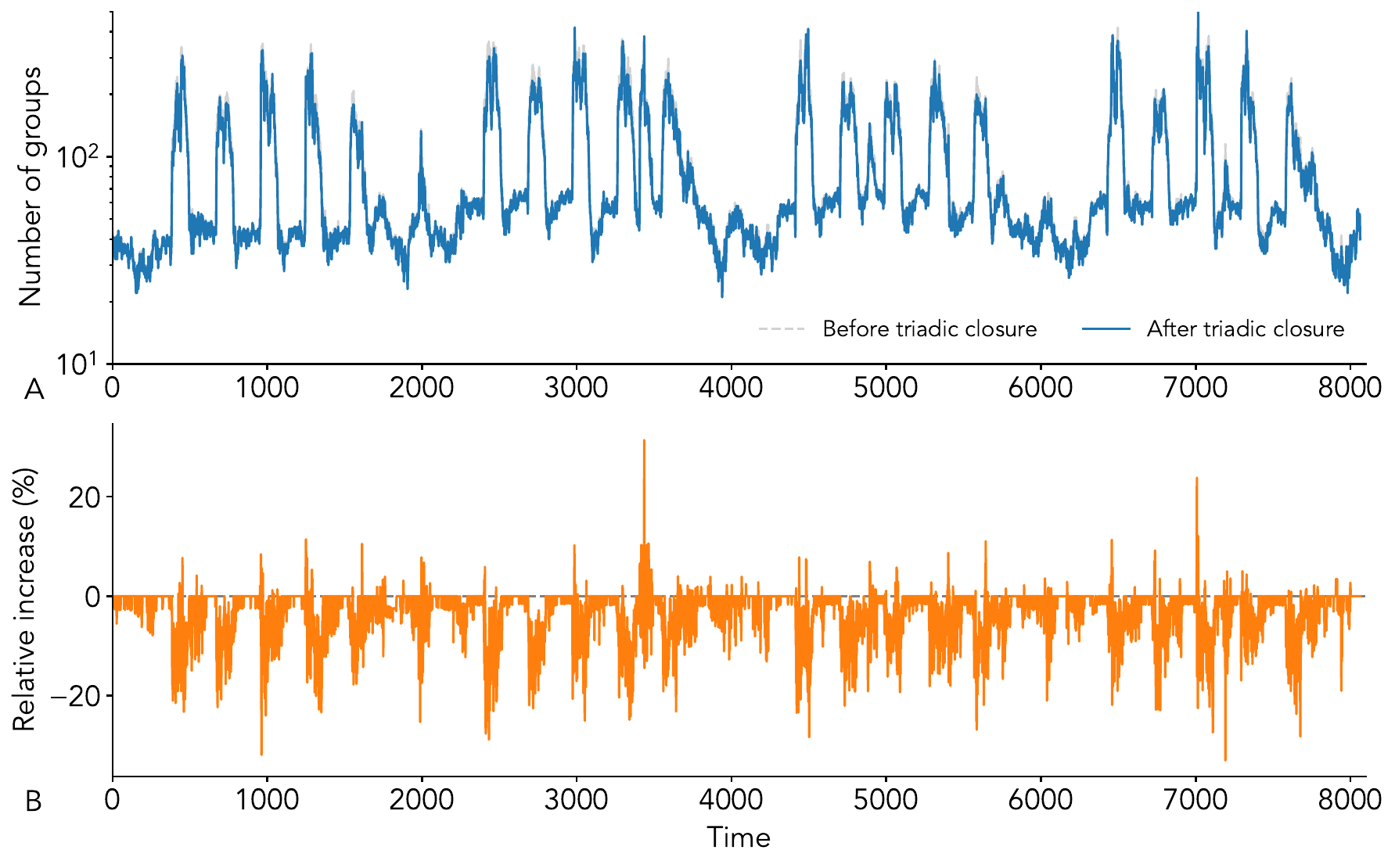}
	\end{center}
	\caption[CNS data: impact of triadic closure on number of groups]{Preprocessing of the CNS dataset: impact of the triadic closure on the number of groups in time ({\it A}), and relative change ({\it B}).
	}\label{fig:SI:CNS:tri-clo_num_groups}
\end{figure}

Third, we perform triadic closure. Namely, if at time $t$ a user $i$ scans a user $j$ and user $j$ scans a user $k$, then we also add a record (if not already present) of an active scan between user $i$ and user $k$. We assign to this interaction the minimum RSSI between the other two.
One of the potential pitfalls of performing triadic closure is the addition of many links to events that have already a low RSSI. In particular, if we filter the number of newly added links ---due to the triadic closure--- by RSSI, we notice that this number scales as a power law with the RSSI [see \SI{Fig.~\ref{fig:SI:CNS:tri-clo_RSSI}}]. In order to avoid closing triangles associated to ``weak'' events, we select an additional threshold of $-75dBm$ for the RSSI of the newly added links. This is chosen as the lowest threshold that preserves the group-size distribution across the different contexts [see \SI{Fig.~\ref{fig:SI:CNS:tri-clo_thresholds}}]. Figures \SI{Fig.~\ref{fig:SI:CNS:tri-clo_num_interactions} and Fig.~\ref{fig:SI:CNS:tri-clo_num_groups}} show the impact of the triadic closure ---with the chosen threshold--- on the number of links and groups in time, respectively. When observed through time, the added links by themselves do not significantly affect the number of tracked links, but help reducing the number of groups ---merging together components that would be disconnected otherwise.

As a final step, we check whether the links removed during the procedure involved were the only interactions of an involved node at that particular time (also considering the triadic closure). In this case, we add back the node to the records and declare it as isolated. We finally end up with a pre-processed data set of 3,991,329 records.

\subsubsection*{DyLNet Study}

The DyLNet data set was collected with the purpose of observing longitudinally the co-evolution of social network and language development of children in pre-school age. The data collection was carried out in a French preschool by recording the proxy social interactions and voice of 174 children between age 3 to 6 and their teachers and assistants. In this study we rely on data openly shared in~\cite{dai2022longitudinal} and focus on the proximity social interaction data that was recorded over 9 sessions (5 morning and 4 afternoon periods ---there are no classes in France on Wednesdays) per week, in 10 consecutive months during a single academic year. The data collection was carried out by using autonomous Radio Frequency Identification (RFID) Wireless Proximity Sensors (employing IEEE 802.15.4 low-rate wireless standard to communicate) installed on participants. Ground-truth data was collected in situ or via controlled experimental settings. Badges broadcasted a ‘hello’ packet with 0 dBm transmission power for 384 $\mu s$ every 5 seconds, otherwise they were in listening mode to record the badge ID and RSSI of other proxy badges if the received signal reached the minimum sensitivity value of -94 dBm. Mutually observed badges were paired to indicate proximity interactions, and were further pre-processed to finally obtain an undirected temporal network~\cite{dai2020temporal}. Interactions in this network indicate face-to-face proximity of participants within 2 meters with temporal resolution of 5 seconds.

Taking the reconstructed network~\cite{dai2022longitudinal} as a starting point, we remove teachers from the data, restricting our attention to children. The data are also enriched with information. For example, the record of each pairwise interaction comes with a 5 digits label that tracks the context category of each of the two individuals at the beginning and end of the interaction. Leveraging this information, as well as the identity and class membership of each individual, as per the CNS, allows us to split interactions into two categories:
\begin{itemize}
	\item {\it in-class}: interactions among children belonging to the same class that starts and finish during class time. Spurious interactions of children belonging to different classes during class time or interactions that start in class and end during the free time are thus removed;
	\item {\it out-of-class}: interactions among children of any class that start and finish during the free time. Spurious interactions that start in class and end during free time or viceversa are thus removed.
\end{itemize}
Differently from the CNS data set, no data collection was performed over the weekends. 
Although the resolution of the original data is 5 seconds, since there was no central unit synchronising the clocks of the badges attached to each participant (they were in-sync only once per a day), we remove interactions that last for less than 10 seconds. Finally, differently from the CNS, the DyLNet data records do not explicitly include isolated participants, i.e., a child that a given timestamp does not participate to any interaction. We thus ``add back'' isolated records for each child in all those timestamps in which that child does not interact with other nodes, but such that the child had at least one interaction during the same school session.

\subsection*{Computing node transition matrices}

The node transition matrices measure the conditional probabilities of moving across groups of different sizes.
Each matrix is constructed ---from a node-centric point of view--- by counting, for each node, the number of observed transitions at consecutive times across two different groups of sizes $k$ and $k^\prime$. Let us denote by  $\sigma_i^t$ the group where node $i$ belongs at time $t$, and by $n_i^t= |\sigma_i^t|$ its size.
Let us consider now a random node $x$ at a time $\tau$ in which it undergoes a group change. We compute the probability that it changes from a group of size $k$ to a group of size $k'$, 
$P(n_x^{\tau+1} =k^{\prime}|n_x^\tau=k)$, as
\begin{equation}\label{eq:transition_matrix}
		T_{kk^{\prime}}\equiv 
  P(n_x^{\tau+1} =k^{\prime}|n_x^\tau=k)
		= \frac{\sum_{t,i}\delta_{n_i^t k}\delta_{n_i^{t+1} k^{\prime}}(1-\delta_{\sigma_i^t\sigma_i^{t+1}})}{k^{\prime}\sum_{k'',t,i}\delta_{n_i^t k}\delta_{n_i^{t+1} k''}(1-\delta_{\sigma_i^t\sigma_i^{t+1}})},
\end{equation}
where the sum at the numerator takes into account all the transitions of all nodes $i\in\mathcal{V}$ taking place at any time $t$, from a group of size $k$ to a group of size $k'$, and the sum at the denominator
takes into account all such transitions, but to a group of any size $k''$ ($\delta_{x,y}$ is the Kronecker delta function, i.e., $\delta_{x,y}=1$ if $x=y$ and zero otherwise). The normalization by the size of the target group in Eq.~\eqref{eq:transition_matrix} ensures that changes to groups of different sizes are comparable. Without this, a single node leaving a group of, say, 5 nodes ---assuming no further changes to the group--- would result in 4 contributions (the remaining nodes) to the transitions from size 5 to size 4.

\subsection*{Computing group aggregation and disaggregation matrices}

Studying group aggregation and disaggregation helps us to understand how groups that form/dismantle behave just before/after the event. Each group interaction $\sigma$, of size $k\equiv|\sigma|$, is associated to a time of birth $t_{\beta}$ and a time of death $t_{\delta}$ defining a temporal span $\tau=t_{\delta}-_{\beta}$ in which all the members of the group stayed continuously together. To study the aggregation and disaggregation phases, we look at how the $k$ members of $\sigma$ were respectively distributed among groups at $t_{\beta}-1$ and $t_{\delta}+1$ (if these timestamps are present within the considered context of interaction). In particular, the probability heatmaps shown in Fig.~\ref{fig:dis-agg} are constructed, for each group size $k$, from the frequencies of the sizes of the maximal sub-groups of $\sigma$ right before its birth,
\begin{equation}\label{eq:group-agg}
	\max\limits_{\sigma^{\prime}\in\mathcal{K}^{t_{\beta}-1}: |\sigma^{\prime}|<|\sigma|} |\sigma \cap \sigma^{\prime}|,
\end{equation}
\noindent and right after its death,
\begin{equation}\label{eq:group-disagg}
	\max\limits_{\sigma^{\prime}\in\mathcal{K}^{t_{\delta}+1}: |\sigma^{\prime}|<|\sigma|} |\sigma \cap \sigma^{\prime}|.
\end{equation}

Notice how we intentionally restrict our attention to the sub-groups of smaller sizes, thus splitting the dynamics into groups that either grow or shrink. Within this dichotomy, for example, a group of size 3 that detaches from a group of size 5 will not contribute to building the probability distribution associated to the aggregation dynamics for $k=3$, but only to the disaggregation one for $k=5$.  

\subsection*{Computing group-change probabilities}
\begin{figure}
	\centering
	\includegraphics[width=\textwidth]{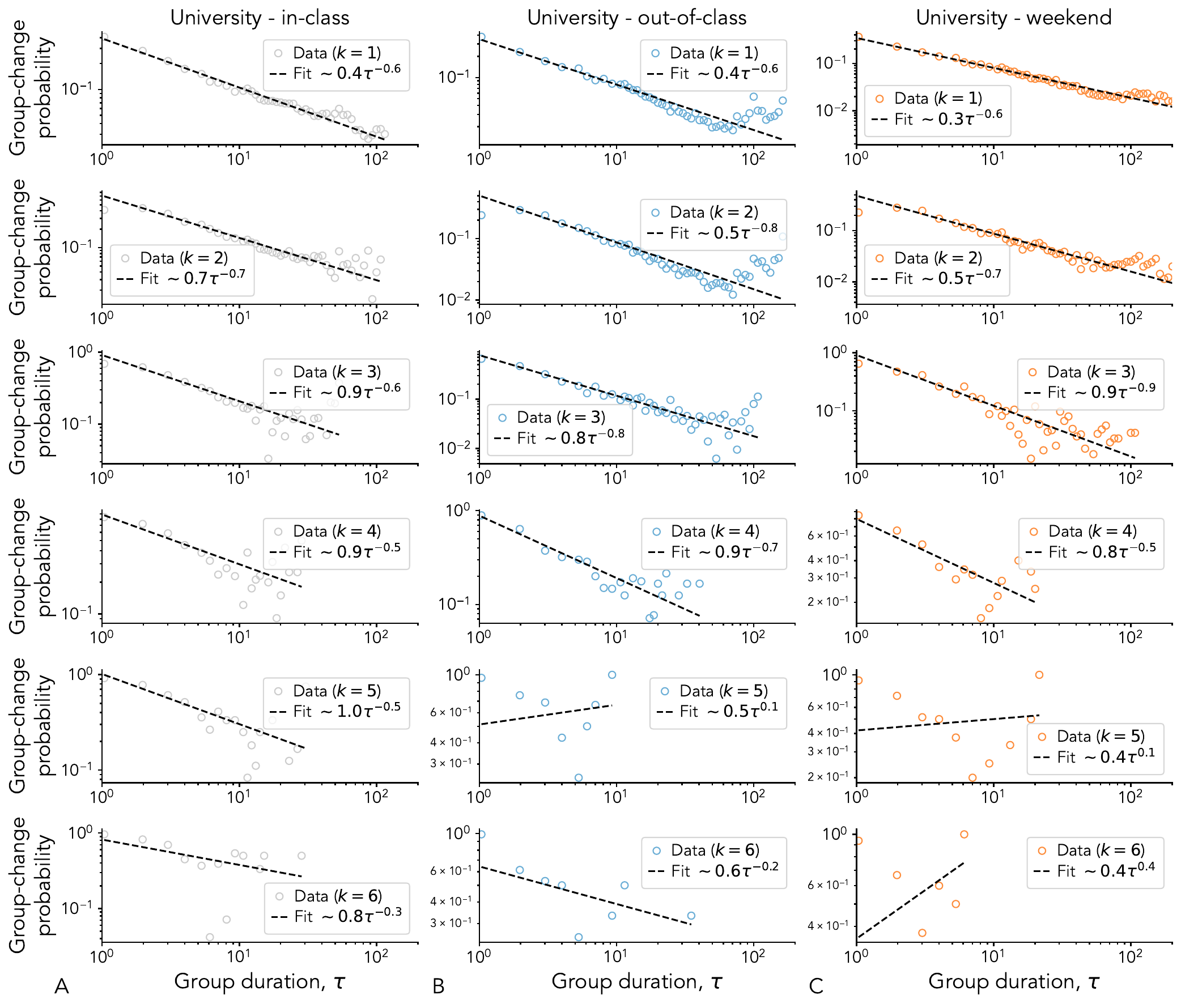}
	\caption[CNS data: group-change probability disaggregated by group size]{Fitting the group-change probability from the CNS data set by group size $k$ and by context of interaction: in-class ({\it A}), out-of-class ({\it B}), and weekend ({\it C}). Each row corresponds to a different size $k$ of the group interaction. 
		We plot the probability for a node belonging to a group of size $k$ to leave it for a different one after exactly $\tau$ timestamps.
		Points are binned empirical results, dashed lines represent a power-law fit of the form $b\tau^{\beta}$ ---values reported in each panel.
	}\label{fig:SI:CNS:group-change_prob_by_size}
\end{figure}

\begin{figure}
	\centering
	\includegraphics[width=.7\textwidth]{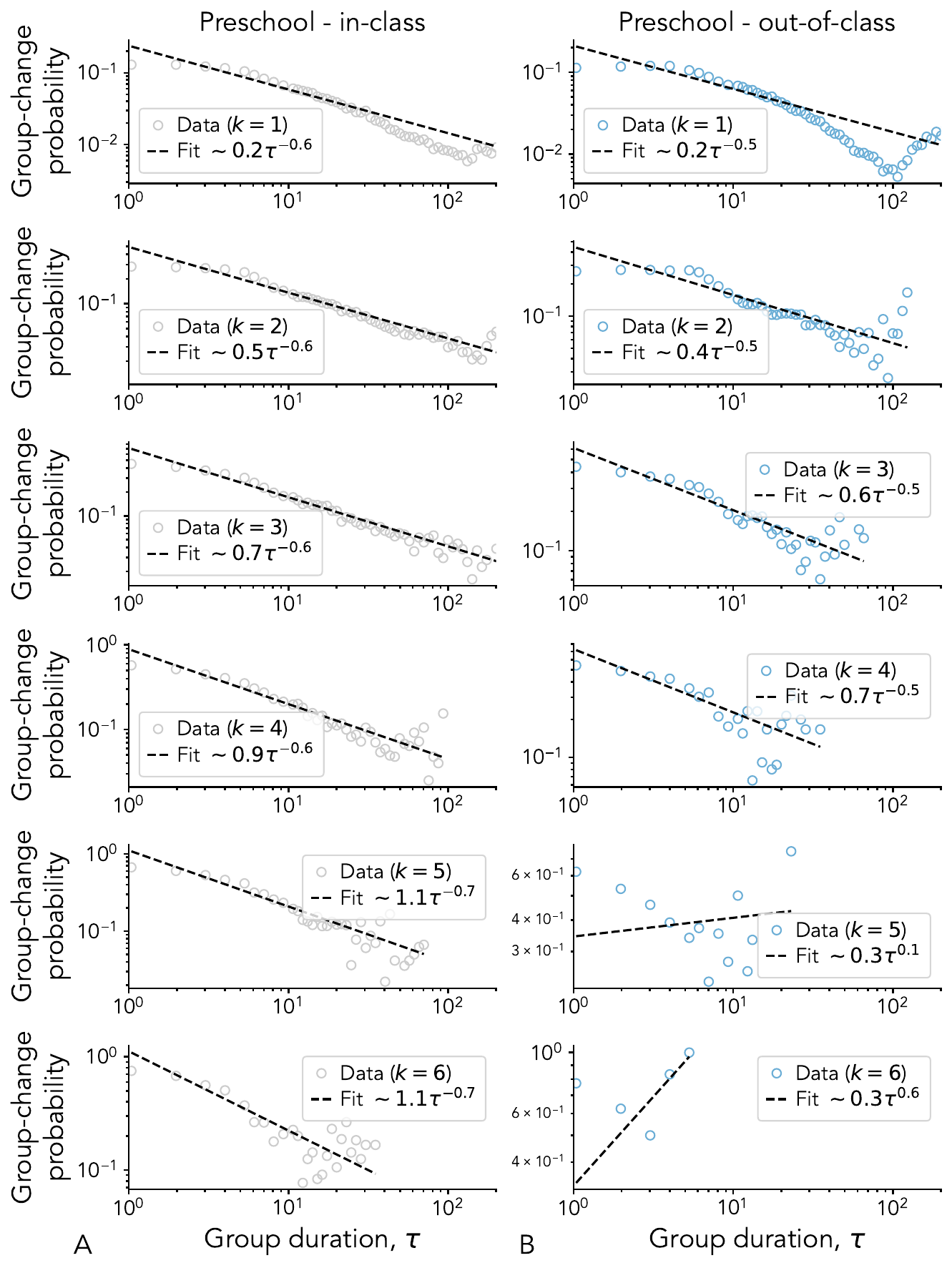}
	\caption[DyLNet data: group-change probability disaggregated by group size]{Fitting the group-change probability from the DyLNet data set by group size $k$ and by context of interaction: in-class ({\it A}) and out-of-class ({\it B}). Each row corresponds to a different size $k$ of the group interaction. 
		We plot the probability for a node belonging to a group of size $k$ to leave it for a different one after exactly $\tau$ timestamps.
		Points are binned empirical results, dashed lines represent a power-law fit of the form $b\tau^{\beta}$ ---values reported in each panel. 
	}\label{fig:SI:DyLNet:group-change_prob_by_size}
\end{figure}

The group-change probability for each data set and context of interaction is computed by considering for all time steps $t$ and all nodes $i$ the number of times each node, belonging to a group of size $k$, leaves the group after $\tau$ timestamps, over all the possible times. This is defined as:
\begin{equation}\label{eq:group-change_prob}
	P^{\curvearrowright}(k, \tau)=
	\frac{
		\sum\limits_{t,i}(1-\delta_{\sigma_i^{t+1}\sigma_i^t})
		\prod\limits_{\Delta=0}^{\tau+1} \delta_{\sigma_i^t\sigma_i^{t-\Delta}}
		(1-\delta_{\sigma_i^t\sigma_i^{t-\tau}})\delta_{n_i^t k}
	}
	{
		\sum\limits_{t,i}
		\prod\limits_{\Delta=0}^{\tau+1} \delta_{\sigma_i^t\sigma_i^{t-\Delta}}
		(1-\delta_{\sigma_i^t\sigma_i^{t-\tau}})\delta_{n_i^t k}
	}.
\end{equation}
Fig.~\ref{eq:logistic} shows the results aggregated over all group sizes, while the full results from Eq.~\eqref{eq:group-change_prob} are given in \SI{Fig.~\ref{fig:SI:CNS:group-change_prob_by_size} and Fig.~\ref{fig:SI:DyLNet:group-change_prob_by_size}} for the CNS and the DyLNet data, respectively.

\subsection*{Null models for group change}
When checking for the presence of {\it social memory} effects in the empirical and in the synthetic data, we also define two null models for comparison. 
Let us consider the case of a group change performed by a node $i$ that switches from group $\omega_i^t$ at time $t$ to a different group $\omega_i^{t+1}$ at time $t+1$. Notice that, despite the splitting of the datasets into different temporal windows (as given by the different contexts of interaction), we do not have problems at the borders as we restrict our attention to group transitions that were actually recorded in the data. 
The first baseline scenario we consider is the case of a random selection, in which $i$ chooses instead a random group $\omega\in \mathcal{K}^{t+1}$ uniformly at random from the set of available groups at $t+1$. Notice that there will always be at least one group to choose from, that is the one found in the data.
As a second baseline scenario we add to this random selection a constraint on the size, such that the group is chosen uniformly at random from the subset of the groups in $\mathcal{K}^{t+1}$ that have the same size of the target group found in the empirical transition. This second type of null model does not work for the case of an empirical transition towards a group of unitary size (a node that becomes isolated). All these transitions are thus discarded from the computation. Notice however that this does not jeopardise the comparison as the computation of the density of known nodes in this transition always leads to a 0 ---ultimately reducing the differences with the null models.

\subsection*{Model parametrization and fitting}
The model is fitted by selecting the best-performing run among different combinations of parameters and with respect to two target observables. In particular, we perform different realisations of the model for different combinations of the parameters $\theta=\{\epsilon, \alpha, n_0\}$ that take values in the intervals $\epsilon\in[1,30]$, $\alpha\in[0.05,0.95]$, $n_0\in[3,14]$, while keeping constant $N=700$ (as the number of students at the university), $\beta=0.8$ (as measured, see Fig.~\ref{fig:group-change}), and for a number of time steps equals to $T=2000$ (notice that each time step involves the activation of every node in a random order). The optimal set of parameters $\theta^*$ is selected based on a joint minimisation of the Kullback–Leibler (KL) divergence $D_{\text{KL}}(\cdot||\cdot)$ with respect to the logarithm of the empirical group-size distribution $\hat{\text{P}}(k)$ and the node transition matrix $\hat{T}_{kk^{\prime}}$:
\begin{equation}
		\theta^*=\argmin_\theta\Big[ \mu D_{\text{KL}}\big(\hat{\text{P}}(k)||P(k|\theta)\big)
		+[1-\mu]D_{\text{KL}}\big(\hat{T}_{kk^{\prime}}||T(kk^{\prime}|\theta)\big)\Big],
\end{equation}
with $\mu=1/2$.

\section*{Data availability}
The Copenhagen Network Study data are available from the original source~\cite{sapiezynski2019interaction} at \href{https://doi.org/10.6084/m9.figshare.7267433}{doi.org/10.6084/m9.figshare.7267433}.
The DyLNet data are available from the original source~\cite{dai2022longitudinal} at \href{https://doi.org/10.7303/syn26560886}{doi.org/10.7303/syn26560886}. The GESIS data are available upon request from the original source~\cite{genois2023combining} at \href{https://doi.org/10.7802/2351}{doi.org/10.7802/2351}.

\section*{Code availability}
The entire analysis was conducted using Python. In particular, the model was coded using the \href{https://xgi.readthedocs.io/en/stable/}{XGI} Python library for {\it compleX Group Interactions}~\cite{Landry2023}.
All scripts and notebooks that support the findings of this study can be found at the Github repository \href{https://github.com/iaciac/temporal-group-interactions}{github.com/iaciac/temporal-group-interactions}~\cite{iacopini2024temporal_code}.

\section*{Acknowledgements}
The authors are thankful for the insightful discussion with Sicheng Dai about the DyLNet data set. I.I. acknowledges support from the James S. McDonnell Foundation $21^{\text{st}}$ Century Science Initiative (\href{https://doi.org/10.37717/2020-1516}{doi.org/10.37717/2020-1516}). A.B. and M.K. acknowledge support from the Agence Nationale de la Recherche (ANR) project DATAREDUX (ANR-19-CE46-0008). M.K. was supported by the CHIST-ERA project SAI: FWF I 5205-N; the SoBigData++ H2020-871042; the EMOMAP CIVICA projects; and the National Laboratory for Health Security (RRF-2.3.1-21-2022-00006).

\section*{Author contributions statement}
I.I., M.K., A.B. designed and conceived the study. I.I. performed the data analysis, the implementation of the model, and the numerical simulations. I.I., M.K., A.B. analyzed and discussed the results. I.I., M.K., A.B. wrote the manuscript.

\section*{Competing interests} The authors declare that they have no competing interests. The funders had no role in study design, data collection and analysis, decision to publish, or preparation of the manuscript.

\section*{Correspondence} Correspondence and requests for materials should be addressed to I.I.~(email: iacopo.iacopini@nulondon.ac.uk).


\begin{thebibliography}{10}
	\urlstyle{rm}
	\expandafter\ifx\csname url\endcsname\relax
	\def\url#1{\texttt{#1}}\fi
	\expandafter\ifx\csname urlprefix\endcsname\relax\def\urlprefix{URL }\fi
	\expandafter\ifx\csname doiprefix\endcsname\relax\def\doiprefix{DOI: }\fi
	\providecommand{\bibinfo}[2]{#2}
	\providecommand{\eprint}[2][]{\url{#2}}
	
	\bibitem{wasserman1994social}
	\bibinfo{author}{Wasserman, S.} \& \bibinfo{author}{Faust, K.}
	\newblock \emph{\bibinfo{title}{Social Network Analysis: Methods and
			Applications}}.
	\newblock Structural Analysis in the Social Sciences
	(\bibinfo{publisher}{Cambridge University Press}, \bibinfo{year}{1994}).
	
	\bibitem{lehmann2007group}
	\bibinfo{author}{Lehmann, J.}, \bibinfo{author}{Korstjens, A.~H.} \&
	\bibinfo{author}{Dunbar, R.~I.}
	\newblock \bibinfo{journal}{\bibinfo{title}{Group size, grooming and social
			cohesion in primates}}.
	\newblock {\emph{\JournalTitle{Anim. Behav.}}} \textbf{\bibinfo{volume}{74}},
	\bibinfo{pages}{1617--1629},
	\doiprefix\url{https://doi.org/10.1016/j.anbehav.2006.10.025}
	(\bibinfo{year}{2007}).
	
	\bibitem{dunbar2018anatomy}
	\bibinfo{author}{Dunbar, R.~I.}
	\newblock \bibinfo{journal}{\bibinfo{title}{The anatomy of friendship}}.
	\newblock {\emph{\JournalTitle{Trends Cogn. Sci.}}}
	\textbf{\bibinfo{volume}{22}}, \bibinfo{pages}{32--51},
	\doiprefix\url{https://doi.org/10.1016/j.tics.2017.10.004}
	(\bibinfo{year}{2018}).
	
	\bibitem{dunbar2020structure}
	\bibinfo{author}{Dunbar, R.}
	\newblock \bibinfo{journal}{\bibinfo{title}{Structure and function in human and
			primate social networks: Implications for diffusion, network stability and
			health}}.
	\newblock {\emph{\JournalTitle{Proc. R. Soc. A}}}
	\textbf{\bibinfo{volume}{476}}, \bibinfo{pages}{20200446},
	\doiprefix\url{https://doi.org/10.1098/rspa.2020.0446}
	(\bibinfo{year}{2020}).
	
	\bibitem{albert2002statistical}
	\bibinfo{author}{Albert, R.} \& \bibinfo{author}{Barab{\'a}si, A.-L.}
	\newblock \bibinfo{journal}{\bibinfo{title}{Statistical mechanics of complex
			networks}}.
	\newblock {\emph{\JournalTitle{Rev. Mod. Phys.}}}
	\textbf{\bibinfo{volume}{74}}, \bibinfo{pages}{47},
	\doiprefix\url{https://doi.org/10.1103/RevModPhys.74.47}
	(\bibinfo{year}{2002}).
	
	\bibitem{newman2003structure}
	\bibinfo{author}{Newman, M.~E.}
	\newblock \bibinfo{journal}{\bibinfo{title}{The structure and function of
			complex networks}}.
	\newblock {\emph{\JournalTitle{SIAM review}}} \textbf{\bibinfo{volume}{45}},
	\bibinfo{pages}{167--256},
	\doiprefix\url{https://doi.org/10.1137/S003614450342480}
	(\bibinfo{year}{2003}).
	
	\bibitem{barrat2008dynamical}
	\bibinfo{author}{Barrat, A.}, \bibinfo{author}{Barth{\'e}lemy, M.} \&
	\bibinfo{author}{Vespignani, A.}
	\newblock \emph{\bibinfo{title}{Dynamical Processes on Complex Networks}}
	(\bibinfo{publisher}{Cambridge University Press}, \bibinfo{year}{2008}).
	
	\bibitem{latora_nicosia_russo_2017}
	\bibinfo{author}{Latora, V.}, \bibinfo{author}{Nicosia, V.} \&
	\bibinfo{author}{Russo, G.}
	\newblock \emph{\bibinfo{title}{Complex Networks: Principles, Methods and
			Applications}}.
	\newblock Complex Networks: Principles, Methods and Applications
	(\bibinfo{publisher}{Cambridge University Press}, \bibinfo{year}{2017}).
	
	\bibitem{vespignani2018twenty}
	\bibinfo{author}{Vespignani, A.}
	\newblock \bibinfo{title}{Twenty years of network science}
	(\bibinfo{year}{2018}).
	
	\bibitem{holme2012temporal}
	\bibinfo{author}{Holme, P.} \& \bibinfo{author}{Saram{\"a}ki, J.}
	\newblock \bibinfo{journal}{\bibinfo{title}{Temporal networks}}.
	\newblock {\emph{\JournalTitle{Phys. Rep.}}} \textbf{\bibinfo{volume}{519}},
	\bibinfo{pages}{97--125},
	\doiprefix\url{https://doi.org/10.1016/j.physrep.2012.03.001}
	(\bibinfo{year}{2012}).
	
	\bibitem{holme2015modern}
	\bibinfo{author}{Holme, P.}
	\newblock \bibinfo{journal}{\bibinfo{title}{Modern temporal network theory: a
			colloquium}}.
	\newblock {\emph{\JournalTitle{Eur. Phys. J. B}}}
	\textbf{\bibinfo{volume}{88}}, \bibinfo{pages}{1--30},
	\doiprefix\url{https://doi.org/10.1140/epjb/e2015-60657-4}
	(\bibinfo{year}{2015}).
	
	\bibitem{battiston2020networks}
	\bibinfo{author}{Battiston, F.} \emph{et~al.}
	\newblock \bibinfo{journal}{\bibinfo{title}{Networks beyond pairwise
			interactions: {{Structure}} and dynamics}}.
	\newblock {\emph{\JournalTitle{Phys. Rep.}}} \textbf{\bibinfo{volume}{874}},
	\bibinfo{pages}{1--92}, \doiprefix\url{10.1016/j.physrep.2020.05.004}
	(\bibinfo{year}{2020}).
	
	\bibitem{battiston2021physics}
	\bibinfo{author}{Battiston, F.} \emph{et~al.}
	\newblock \bibinfo{journal}{\bibinfo{title}{The physics of higher-order
			interactions in complex systems}}.
	\newblock {\emph{\JournalTitle{Nat. Phys.}}} \textbf{\bibinfo{volume}{17}},
	\bibinfo{pages}{1093--1098}, \doiprefix\url{10.1038/s41567-021-01371-4}
	(\bibinfo{year}{2021}).
	
	\bibitem{torres2021and}
	\bibinfo{author}{Torres, L.}, \bibinfo{author}{Blevins, A.~S.},
	\bibinfo{author}{Bassett, D.} \& \bibinfo{author}{Eliassi-Rad, T.}
	\newblock \bibinfo{journal}{\bibinfo{title}{The why, how, and when of
			representations for complex systems}}.
	\newblock {\emph{\JournalTitle{SIAM Rev.}}} \textbf{\bibinfo{volume}{63}},
	\bibinfo{pages}{435--485}, \doiprefix\url{https://doi.org/10.1137/20M1355896}
	(\bibinfo{year}{2021}).
	
	\bibitem{bianconi_2021}
	\bibinfo{author}{Bianconi, G.}
	\newblock \emph{\bibinfo{title}{Higher-Order Networks}}.
	\newblock Elements in Structure and Dynamics of Complex Networks
	(\bibinfo{publisher}{Cambridge University Press}, \bibinfo{year}{2021}).
	
	\bibitem{milojevic2014principles}
	\bibinfo{author}{Milojevi{\'c}, S.}
	\newblock \bibinfo{journal}{\bibinfo{title}{Principles of scientific research
			team formation and evolution}}.
	\newblock {\emph{\JournalTitle{Proc. Natl. Acad. Sci. U.S.A.}}}
	\textbf{\bibinfo{volume}{111}}, \bibinfo{pages}{3984--3989},
	\doiprefix\url{https://doi.org/10.1073/pnas.1309723111}
	(\bibinfo{year}{2014}).
	
	\bibitem{juul2024hypergraph}
	\bibinfo{author}{Juul, J.~L.}, \bibinfo{author}{Benson, A.~R.} \&
	\bibinfo{author}{Kleinberg, J.}
	\newblock \bibinfo{journal}{\bibinfo{title}{Hypergraph patterns and
			collaboration structure}}.
	\newblock {\emph{\JournalTitle{Front. Phys.}}} \textbf{\bibinfo{volume}{11}},
	\bibinfo{pages}{1301994},
	\doiprefix\url{https://doi.org/10.3389/fphy.2023.1301994}
	(\bibinfo{year}{2024}).
	
	\bibitem{katz2011inferring}
	\bibinfo{author}{Katz, Y.}, \bibinfo{author}{Tunstr{\o}m, K.},
	\bibinfo{author}{Ioannou, C.~C.}, \bibinfo{author}{Huepe, C.} \&
	\bibinfo{author}{Couzin, I.~D.}
	\newblock \bibinfo{journal}{\bibinfo{title}{Inferring the structure and
			dynamics of interactions in schooling fish}}.
	\newblock {\emph{\JournalTitle{Proc. Natl. Acad. Sci. U.S.A.}}}
	\textbf{\bibinfo{volume}{108}}, \bibinfo{pages}{18720--18725},
	\doiprefix\url{https://doi.org/10.1073/pnas.110758310}
	(\bibinfo{year}{2011}).
	
	\bibitem{mcgrath1984groups}
	\bibinfo{author}{McGrath, J.}
	\newblock \emph{\bibinfo{title}{Groups: Interaction and Performance}}
	(\bibinfo{publisher}{Prentice-Hall}, \bibinfo{year}{1984}).
	
	\bibitem{patania2017shape}
	\bibinfo{author}{Patania, A.}, \bibinfo{author}{Petri, G.} \&
	\bibinfo{author}{Vaccarino, F.}
	\newblock \bibinfo{journal}{\bibinfo{title}{The shape of collaborations}}.
	\newblock {\emph{\JournalTitle{EPJ Data Sci.}}} \textbf{\bibinfo{volume}{6}},
	\bibinfo{pages}{1--16},
	\doiprefix\url{https://doi.org/10.1140/epjds/s13688-017-0114-8}
	(\bibinfo{year}{2017}).
	
	\bibitem{benson2018simplicial}
	\bibinfo{author}{Benson, A.~R.}, \bibinfo{author}{Abebe, R.},
	\bibinfo{author}{Schaub, M.~T.}, \bibinfo{author}{Jadbabaie, A.} \&
	\bibinfo{author}{Kleinberg, J.}
	\newblock \bibinfo{journal}{\bibinfo{title}{Simplicial closure and higher-order
			link prediction}}.
	\newblock {\emph{\JournalTitle{Proc. Natl. Acad. Sci. U.S.A.}}}
	\textbf{\bibinfo{volume}{115}}, \bibinfo{pages}{E11221--E11230},
	\doiprefix\url{https://doi.org/10.1073/pnas.1800683115}
	(\bibinfo{year}{2018}).
	
	\bibitem{cencetti2021temporal}
	\bibinfo{author}{Cencetti, G.}, \bibinfo{author}{Battiston, F.},
	\bibinfo{author}{Lepri, B.} \& \bibinfo{author}{Karsai, M.}
	\newblock \bibinfo{journal}{\bibinfo{title}{Temporal properties of higher-order
			interactions in social networks}}.
	\newblock {\emph{\JournalTitle{Sci. Rep.}}} \textbf{\bibinfo{volume}{11}},
	\bibinfo{pages}{1--10},
	\doiprefix\url{https://doi.org/10.1038/s41598-021-86469-8}
	(\bibinfo{year}{2021}).
	
	\bibitem{iacopini2022group}
	\bibinfo{author}{Iacopini, I.}, \bibinfo{author}{Petri, G.},
	\bibinfo{author}{Baronchelli, A.} \& \bibinfo{author}{Barrat, A.}
	\newblock \bibinfo{journal}{\bibinfo{title}{Group interactions modulate
			critical mass dynamics in social convention}}.
	\newblock {\emph{\JournalTitle{Commun., Phys.}}} \textbf{\bibinfo{volume}{5}},
	\bibinfo{pages}{1--10},
	\doiprefix\url{https://doi.org/10.1038/s42005-022-00845-y}
	(\bibinfo{year}{2022}).
	
	\bibitem{korbel2023homophily}
	\bibinfo{author}{Korbel, J.}, \bibinfo{author}{Lindner, S.~D.},
	\bibinfo{author}{Pham, T.~M.}, \bibinfo{author}{Hanel, R.} \&
	\bibinfo{author}{Thurner, S.}
	\newblock \bibinfo{journal}{\bibinfo{title}{Homophily-based social group
			formation in a spin glass self-assembly framework}}.
	\newblock {\emph{\JournalTitle{Phys. Rev. Lett.}}}
	\textbf{\bibinfo{volume}{130}}, \bibinfo{pages}{057401},
	\doiprefix\url{10.1103/PhysRevLett.130.057401} (\bibinfo{year}{2023}).
	
	\bibitem{forsyth2018group}
	\bibinfo{author}{Forsyth, D.~R.}
	\newblock \emph{\bibinfo{title}{Group dynamics}} (\bibinfo{publisher}{Cengage
		Learning}, \bibinfo{year}{2018}).
	
	\bibitem{geard2010competition}
	\bibinfo{author}{Geard, N.} \& \bibinfo{author}{Bullock, S.}
	\newblock \bibinfo{journal}{\bibinfo{title}{Competition and the dynamics of
			group affiliation}}.
	\newblock {\emph{\JournalTitle{Adv. Complex Syst.}}}
	\textbf{\bibinfo{volume}{13}}, \bibinfo{pages}{501--517},
	\doiprefix\url{https://doi.org/10.1142/S0219525910002712}
	(\bibinfo{year}{2010}).
	
	\bibitem{lotito2022higher}
	\bibinfo{author}{Lotito, Q.~F.}, \bibinfo{author}{Musciotto, F.},
	\bibinfo{author}{Montresor, A.} \& \bibinfo{author}{Battiston, F.}
	\newblock \bibinfo{journal}{\bibinfo{title}{Higher-order motif analysis in
			hypergraphs}}.
	\newblock {\emph{\JournalTitle{Commun. Phys.}}} \textbf{\bibinfo{volume}{5}},
	\bibinfo{pages}{1--8},
	\doiprefix\url{https://doi.org/10.1038/s42005-022-00858-7}
	(\bibinfo{year}{2022}).
	
	\bibitem{mancastroppa2023hyper}
	\bibinfo{author}{Mancastroppa, M.}, \bibinfo{author}{Iacopini, I.},
	\bibinfo{author}{Petri, G.} \& \bibinfo{author}{Barrat, A.}
	\newblock \bibinfo{journal}{\bibinfo{title}{Hyper-cores promote localization
			and efficient seeding in higher-order processes}}.
	\newblock {\emph{\JournalTitle{Nat. Commun.}}} \textbf{\bibinfo{volume}{14}},
	\bibinfo{pages}{6223},
	\doiprefix\url{https://doi.org/10.1038/s41467-023-41887-2}
	(\bibinfo{year}{2023}).
	
	\bibitem{zhao2011social}
	\bibinfo{author}{Zhao, K.}, \bibinfo{author}{Stehl{\'e}, J.},
	\bibinfo{author}{Bianconi, G.} \& \bibinfo{author}{Barrat, A.}
	\newblock \bibinfo{journal}{\bibinfo{title}{Social network dynamics of
			face-to-face interactions}}.
	\newblock {\emph{\JournalTitle{Phys. Rev. E}}} \textbf{\bibinfo{volume}{83}},
	\bibinfo{pages}{056109}, \doiprefix\url{10.1103/PhysRevE.83.056109}
	(\bibinfo{year}{2011}).
	
	\bibitem{ceria2023temporal}
	\bibinfo{author}{Ceria, A.} \& \bibinfo{author}{Wang, H.}
	\newblock \bibinfo{journal}{\bibinfo{title}{Temporal-topological properties of
			higher-order evolving networks}}.
	\newblock {\emph{\JournalTitle{Sci. Rep.}}} \textbf{\bibinfo{volume}{13}},
	\doiprefix\url{https://doi.org/10.1038/s41598-023-32253-9}
	(\bibinfo{year}{2023}).
	
	\bibitem{gallo2024higher}
	\bibinfo{author}{Gallo, L.}, \bibinfo{author}{Lacasa, L.},
	\bibinfo{author}{Latora, V.} \& \bibinfo{author}{Battiston, F.}
	\newblock \bibinfo{journal}{\bibinfo{title}{Higher-order correlations reveal
			complex memory in temporal hypergraphs}}.
	\newblock {\emph{\JournalTitle{Nat. Commun.}}} \textbf{\bibinfo{volume}{15}},
	\bibinfo{pages}{4754},
	\doiprefix\url{https://doi.org/10.1038/s41467-024-48578-6}
	(\bibinfo{year}{2024}).
	
	\bibitem{sekara2016fundamental}
	\bibinfo{author}{Sekara, V.}, \bibinfo{author}{Stopczynski, A.} \&
	\bibinfo{author}{Lehmann, S.}
	\newblock \bibinfo{journal}{\bibinfo{title}{Fundamental structures of dynamic
			social networks}}.
	\newblock {\emph{\JournalTitle{Proc. Natl. Acad. Sci. U.S.A.}}}
	\textbf{\bibinfo{volume}{113}}, \bibinfo{pages}{9977--9982},
	\doiprefix\url{https://doi.org/10.1073/pnas.1602803113}
	(\bibinfo{year}{2016}).
	
	\bibitem{castellano2009statistical}
	\bibinfo{author}{Castellano, C.}, \bibinfo{author}{Fortunato, S.} \&
	\bibinfo{author}{Loreto, V.}
	\newblock \bibinfo{journal}{\bibinfo{title}{Statistical physics of social
			dynamics}}.
	\newblock {\emph{\JournalTitle{Rev. Mod. Phys.}}}
	\textbf{\bibinfo{volume}{81}}, \bibinfo{pages}{591},
	\doiprefix\url{https://doi.org/10.1103/RevModPhys.81.591}
	(\bibinfo{year}{2009}).
	
	\bibitem{vespignani2012modelling}
	\bibinfo{author}{Vespignani, A.}
	\newblock \bibinfo{journal}{\bibinfo{title}{Modelling dynamical processes in
			complex socio-technical systems}}.
	\newblock {\emph{\JournalTitle{Nat. Phys.}}} \textbf{\bibinfo{volume}{8}},
	\bibinfo{pages}{32--39}, \doiprefix\url{https://doi.org/10.1038/nphys2160}
	(\bibinfo{year}{2012}).
	
	\bibitem{pastor2015epidemic}
	\bibinfo{author}{Pastor-Satorras, R.}, \bibinfo{author}{Castellano, C.},
	\bibinfo{author}{Van~Mieghem, P.} \& \bibinfo{author}{Vespignani, A.}
	\newblock \bibinfo{journal}{\bibinfo{title}{Epidemic processes in complex
			networks}}.
	\newblock {\emph{\JournalTitle{Rev. Mod. Phys.}}}
	\textbf{\bibinfo{volume}{87}}, \bibinfo{pages}{925--979},
	\doiprefix\url{10.1103/RevModPhys.87.925} (\bibinfo{year}{2015}).
	
	\bibitem{vestergaard2014memory}
	\bibinfo{author}{Vestergaard, C.~L.}, \bibinfo{author}{G{\'e}nois, M.} \&
	\bibinfo{author}{Barrat, A.}
	\newblock \bibinfo{journal}{\bibinfo{title}{How memory generates heterogeneous
			dynamics in temporal networks}}.
	\newblock {\emph{\JournalTitle{Phys. Rev. E}}} \textbf{\bibinfo{volume}{90}},
	\bibinfo{pages}{042805},
	\doiprefix\url{https://doi.org/10.1103/PhysRevE.90.042805}
	(\bibinfo{year}{2014}).
	
	\bibitem{iacopini2019simplicial}
	\bibinfo{author}{Iacopini, I.}, \bibinfo{author}{Petri, G.},
	\bibinfo{author}{Barrat, A.} \& \bibinfo{author}{Latora, V.}
	\newblock \bibinfo{journal}{\bibinfo{title}{Simplicial models of social
			contagion}}.
	\newblock {\emph{\JournalTitle{Nat. Commun.}}} \textbf{\bibinfo{volume}{10}},
	\bibinfo{pages}{2485},
	\doiprefix\url{https://doi.org/10.1038/s41467-019-10431-6}
	(\bibinfo{year}{2019}).
	
	\bibitem{st2022influential}
	\bibinfo{author}{St-Onge, G.} \emph{et~al.}
	\newblock \bibinfo{journal}{\bibinfo{title}{Influential groups for seeding and
			sustaining nonlinear contagion in heterogeneous hypergraphs}}.
	\newblock {\emph{\JournalTitle{Commun. Phys.}}} \textbf{\bibinfo{volume}{5}},
	\bibinfo{pages}{1--16},
	\doiprefix\url{https://doi.org/10.1038/s42005-021-00788-w}
	(\bibinfo{year}{2022}).
	
	\bibitem{papanikolaou2022consensus}
	\bibinfo{author}{Papanikolaou, N.}, \bibinfo{author}{Vaccario, G.},
	\bibinfo{author}{Hormann, E.}, \bibinfo{author}{Lambiotte, R.} \&
	\bibinfo{author}{Schweitzer, F.}
	\newblock \bibinfo{journal}{\bibinfo{title}{Consensus from group interactions:
			An adaptive voter model on hypergraphs}}.
	\newblock {\emph{\JournalTitle{Phys. Rev. E}}} \textbf{\bibinfo{volume}{105}},
	\bibinfo{pages}{054307}, \doiprefix\url{10.1103/PhysRevE.105.054307}
	(\bibinfo{year}{2022}).
	
	\bibitem{sheng2024strategy}
	\bibinfo{author}{Sheng, A.}, \bibinfo{author}{Su, Q.}, \bibinfo{author}{Wang,
		L.} \& \bibinfo{author}{Plotkin, J.~B.}
	\newblock \bibinfo{journal}{\bibinfo{title}{Strategy evolution on higher-order
			networks}}.
	\newblock {\emph{\JournalTitle{Nat. Comput. Sci.}}} \bibinfo{pages}{1--11},
	\doiprefix\url{https://doi.org/10.1038/s43588-024-00621-8}
	(\bibinfo{year}{2024}).
	
	\bibitem{civilini2024explosive}
	\bibinfo{author}{Civilini, A.}, \bibinfo{author}{Sadekar, O.},
	\bibinfo{author}{Battiston, F.}, \bibinfo{author}{G{\'o}mez-Garde{\~n}es, J.}
	\& \bibinfo{author}{Latora, V.}
	\newblock \bibinfo{journal}{\bibinfo{title}{Explosive cooperation in social
			dilemmas on higher-order networks}}.
	\newblock {\emph{\JournalTitle{Phys. Rev. Lett.}}}
	\textbf{\bibinfo{volume}{132}}, \bibinfo{pages}{167401},
	\doiprefix\url{https://doi.org/10.1103/PhysRevLett.132.167401}
	(\bibinfo{year}{2024}).
	
	\bibitem{Chowdhary_2021}
	\bibinfo{author}{Chowdhary, S.}, \bibinfo{author}{Kumar, A.},
	\bibinfo{author}{Cencetti, G.}, \bibinfo{author}{Iacopini, I.} \&
	\bibinfo{author}{Battiston, F.}
	\newblock \bibinfo{journal}{\bibinfo{title}{Simplicial contagion in temporal
			higher-order networks}}.
	\newblock {\emph{\JournalTitle{J. Phys. Complexity}}}
	\textbf{\bibinfo{volume}{2}}, \bibinfo{pages}{035019},
	\doiprefix\url{https://doi.org/10.1088/2632-072X/ac12bd}
	(\bibinfo{year}{2021}).
	
	\bibitem{sapiezynski2019interaction}
	\bibinfo{author}{Sapiezynski, P.}, \bibinfo{author}{Stopczynski, A.},
	\bibinfo{author}{Lassen, D.~D.} \& \bibinfo{author}{Lehmann, S.}
	\newblock \bibinfo{journal}{\bibinfo{title}{Interaction data from the
			copenhagen networks study}}.
	\newblock {\emph{\JournalTitle{Sci. Data}}} \textbf{\bibinfo{volume}{6}},
	\bibinfo{pages}{1--10},
	\doiprefix\url{https://doi.org/10.1038/s41597-019-0325-x}
	(\bibinfo{year}{2019}).
	
	\bibitem{dai2022longitudinal}
	\bibinfo{author}{Dai, S.} \emph{et~al.}
	\newblock \bibinfo{journal}{\bibinfo{title}{Longitudinal data collection to
			follow social network and language development dynamics at preschool}}.
	\newblock {\emph{\JournalTitle{Sci. Data}}} \textbf{\bibinfo{volume}{9}},
	\bibinfo{pages}{1--17},
	\doiprefix\url{https://doi.org/10.1038/s41597-022-01756-x}
	(\bibinfo{year}{2022}).
	
	\bibitem{dai2020temporal}
	\bibinfo{author}{Dai, S.} \emph{et~al.}
	\newblock \bibinfo{journal}{\bibinfo{title}{Temporal social network
			reconstruction using wireless proximity sensors: model selection and
			consequences}}.
	\newblock {\emph{\JournalTitle{EPJ Data Sci.}}} \textbf{\bibinfo{volume}{9}},
	\bibinfo{pages}{19},
	\doiprefix\url{https://doi.org/10.1140/epjds/s13688-020-00237-8}
	(\bibinfo{year}{2020}).
	
	\bibitem{braha2006centrality}
	\bibinfo{author}{Braha, D.} \& \bibinfo{author}{Bar-Yam, Y.}
	\newblock \bibinfo{journal}{\bibinfo{title}{From centrality to temporary fame:
			Dynamic centrality in complex networks}}.
	\newblock {\emph{\JournalTitle{Complexity}}} \textbf{\bibinfo{volume}{12}},
	\bibinfo{pages}{59--63}, \doiprefix\url{https://doi.org/10.1002/cplx.20156}
	(\bibinfo{year}{2006}).
	
	\bibitem{braha2009time}
	\bibinfo{author}{Braha, D.} \& \bibinfo{author}{Bar-Yam, Y.}
	\newblock \bibinfo{title}{Time-dependent complex networks: Dynamic centrality,
		dynamic motifs, and cycles of social interactions}.
	\newblock In \emph{\bibinfo{booktitle}{Adaptive networks: Theory, models and
			applications}}, \bibinfo{pages}{39--50},
	\doiprefix\url{https://doi.org/10.1007/978-3-642-01284-6_3}
	(\bibinfo{publisher}{Springer}, \bibinfo{year}{2009}).
	
	\bibitem{pedreschi2020dynamic}
	\bibinfo{author}{Pedreschi, N.} \emph{et~al.}
	\newblock \bibinfo{journal}{\bibinfo{title}{Dynamic core-periphery structure of
			information sharing networks in entorhinal cortex and hippocampus}}.
	\newblock {\emph{\JournalTitle{Netw. Neurosci.}}} \textbf{\bibinfo{volume}{4}},
	\bibinfo{pages}{946--975},
	\doiprefix\url{https://doi.org/10.1162/netn_a_00142} (\bibinfo{year}{2020}).
	
	\bibitem{alstott2014powerlaw}
	\bibinfo{author}{Alstott, J.}, \bibinfo{author}{Bullmore, E.} \&
	\bibinfo{author}{Plenz, D.}
	\newblock \bibinfo{journal}{\bibinfo{title}{powerlaw: a python package for
			analysis of heavy-tailed distributions}}.
	\newblock {\emph{\JournalTitle{PLoS One}}} \textbf{\bibinfo{volume}{9}},
	\bibinfo{pages}{e85777},
	\doiprefix\url{https://doi.org/10.1371/journal.pone.0085777}
	(\bibinfo{year}{2014}).
	
	\bibitem{cattuto2010dynamics}
	\bibinfo{author}{Cattuto, C.} \emph{et~al.}
	\newblock \bibinfo{journal}{\bibinfo{title}{Dynamics of person-to-person
			interactions from distributed rfid sensor networks}}.
	\newblock {\emph{\JournalTitle{PLoS One}}} \textbf{\bibinfo{volume}{5}},
	\bibinfo{pages}{e11596},
	\doiprefix\url{https://doi.org/10.1371/journal.pone.0011596}
	(\bibinfo{year}{2010}).
	
	\bibitem{barrat2014measuring}
	\bibinfo{author}{Barrat, A.}, \bibinfo{author}{Cattuto, C.},
	\bibinfo{author}{Tozzi, A.~E.}, \bibinfo{author}{Vanhems, P.} \&
	\bibinfo{author}{Voirin, N.}
	\newblock \bibinfo{journal}{\bibinfo{title}{Measuring contact patterns with
			wearable sensors: methods, data characteristics and applications to
			data-driven simulations of infectious diseases}}.
	\newblock {\emph{\JournalTitle{Clin. Microbiol. Infect.}}}
	\textbf{\bibinfo{volume}{20}}, \bibinfo{pages}{10--16},
	\doiprefix\url{https://doi.org/10.1111/1469-0691.12472}
	(\bibinfo{year}{2014}).
	
	\bibitem{perra2012activity}
	\bibinfo{author}{Perra, N.}, \bibinfo{author}{Gon{\c{c}}alves, B.},
	\bibinfo{author}{Pastor-Satorras, R.} \& \bibinfo{author}{Vespignani, A.}
	\newblock \bibinfo{journal}{\bibinfo{title}{Activity driven modeling of time
			varying networks}}.
	\newblock {\emph{\JournalTitle{Sci. Rep.}}} \textbf{\bibinfo{volume}{2}},
	\bibinfo{pages}{1--7}, \doiprefix\url{https://doi.org/10.1038/srep00469}
	(\bibinfo{year}{2012}).
	
	\bibitem{starnini2013modeling}
	\bibinfo{author}{Starnini, M.}, \bibinfo{author}{Baronchelli, A.} \&
	\bibinfo{author}{Pastor-Satorras, R.}
	\newblock \bibinfo{journal}{\bibinfo{title}{Modeling human dynamics of
			face-to-face interaction networks}}.
	\newblock {\emph{\JournalTitle{Phys. Rev. Lett.}}}
	\textbf{\bibinfo{volume}{110}}, \bibinfo{pages}{168701},
	\doiprefix\url{10.1103/PhysRevLett.110.168701} (\bibinfo{year}{2013}).
	
	\bibitem{karsai2014time}
	\bibinfo{author}{Karsai, M.}, \bibinfo{author}{Perra, N.} \&
	\bibinfo{author}{Vespignani, A.}
	\newblock \bibinfo{journal}{\bibinfo{title}{Time varying networks and the
			weakness of strong ties}}.
	\newblock {\emph{\JournalTitle{Sci. Rep.}}} \textbf{\bibinfo{volume}{4}},
	\bibinfo{pages}{1--7}, \doiprefix\url{https://doi.org/10.1038/srep04001}
	(\bibinfo{year}{2014}).
	
	\bibitem{nadini2018epidemic}
	\bibinfo{author}{Nadini, M.} \emph{et~al.}
	\newblock \bibinfo{journal}{\bibinfo{title}{Epidemic spreading in modular
			time-varying networks}}.
	\newblock {\emph{\JournalTitle{Sci. Rep.}}} \textbf{\bibinfo{volume}{8}},
	\bibinfo{pages}{1--11},
	\doiprefix\url{https://doi.org/10.1038/s41598-018-20908-x}
	(\bibinfo{year}{2018}).
	
	\bibitem{lebail2023modelling}
	\bibinfo{author}{Le~Bail, D.}, \bibinfo{author}{G\'enois, M.} \&
	\bibinfo{author}{Barrat, A.}
	\newblock \bibinfo{journal}{\bibinfo{title}{Modeling framework unifying contact
			and social networks}}.
	\newblock {\emph{\JournalTitle{Phys. Rev. E}}} \textbf{\bibinfo{volume}{107}},
	\bibinfo{pages}{024301}, \doiprefix\url{10.1103/PhysRevE.107.024301}
	(\bibinfo{year}{2023}).
	
	\bibitem{stehle2010dynamical}
	\bibinfo{author}{Stehl{\'e}, J.}, \bibinfo{author}{Barrat, A.} \&
	\bibinfo{author}{Bianconi, G.}
	\newblock \bibinfo{journal}{\bibinfo{title}{Dynamical and bursty interactions
			in social networks}}.
	\newblock {\emph{\JournalTitle{Phys. Rev. E}}} \textbf{\bibinfo{volume}{81}},
	\bibinfo{pages}{035101}, \doiprefix\url{10.1103/PhysRevE.81.035101}
	(\bibinfo{year}{2010}).
	
	\bibitem{petri2018simplicial}
	\bibinfo{author}{Petri, G.} \& \bibinfo{author}{Barrat, A.}
	\newblock \bibinfo{journal}{\bibinfo{title}{Simplicial activity driven model}}.
	\newblock {\emph{\JournalTitle{Phys. Rev. Lett.}}}
	\textbf{\bibinfo{volume}{121}}, \bibinfo{pages}{228301},
	\doiprefix\url{10.1103/PhysRevLett.121.228301} (\bibinfo{year}{2018}).
	
	\bibitem{hatcher2002algebraic}
	\bibinfo{author}{Hatcher, A.}, \bibinfo{author}{Press, C.~U.} \&
	\bibinfo{author}{of~Mathematics, C. U.~D.}
	\newblock \emph{\bibinfo{title}{Algebraic Topology}}.
	\newblock Algebraic Topology (\bibinfo{publisher}{Cambridge University Press},
	\bibinfo{year}{2002}).
	
	\bibitem{simmel1902number}
	\bibinfo{author}{Simmel, G.}
	\newblock \bibinfo{journal}{\bibinfo{title}{The number of members as
			determining the sociological form of the group}}.
	\newblock {\emph{\JournalTitle{Am. J. Sociol.}}} \textbf{\bibinfo{volume}{8}},
	\bibinfo{pages}{1--46} (\bibinfo{year}{1902}).
	
	\bibitem{egbert1997schisming}
	\bibinfo{author}{Egbert, M.~M.}
	\newblock \bibinfo{journal}{\bibinfo{title}{Schisming: The collaborative
			transformation from a single conversation to multiple conversations}}.
	\newblock {\emph{\JournalTitle{Res. Lang. Soc.}}}
	\textbf{\bibinfo{volume}{30}}, \bibinfo{pages}{1--51},
	\doiprefix\url{https://doi.org/10.1207/s15327973rlsi3001_1}
	(\bibinfo{year}{1997}).
	
	\bibitem{genois2023combining}
	\bibinfo{author}{G{\'e}nois, M.} \emph{et~al.}
	\newblock \bibinfo{journal}{\bibinfo{title}{Combining sensors and surveys to
			study social interactions: A case of four science conferences}}.
	\newblock {\emph{\JournalTitle{Pers. Sci.}}} \textbf{\bibinfo{volume}{4}},
	\bibinfo{pages}{1--24}, \doiprefix\url{https://doi.org/10.5964/ps.9957}
	(\bibinfo{year}{2023}).
	
	\bibitem{shang2023non}
	\bibinfo{author}{Shang, Y.}
	\newblock \bibinfo{journal}{\bibinfo{title}{Non-linear consensus dynamics on
			temporal hypergraphs with random noisy higher-order interactions}}.
	\newblock {\emph{\JournalTitle{J. Complex Netw.}}}
	\textbf{\bibinfo{volume}{11}}, \bibinfo{pages}{cnad009},
	\doiprefix\url{https://doi.org/10.1093/comnet/cnad009}
	(\bibinfo{year}{2023}).
	
	\bibitem{barrat2022social}
	\bibinfo{author}{Barrat, A.}, \bibinfo{author}{Ferraz~de Arruda, G.},
	\bibinfo{author}{Iacopini, I.} \& \bibinfo{author}{Moreno, Y.}
	\newblock \bibinfo{title}{Social contagion on higher-order structures}.
	\newblock In \emph{\bibinfo{booktitle}{Higher-Order Systems}},
	\bibinfo{pages}{329--346},
	\doiprefix\url{https://10.1007/978-3-030-91374-8_13}
	(\bibinfo{publisher}{Springer}, \bibinfo{year}{2022}).
	
	\bibitem{neuhauser2022consensus}
	\bibinfo{author}{Neuh{\"a}user, L.}, \bibinfo{author}{Lambiotte, R.} \&
	\bibinfo{author}{Schaub, M.~T.}
	\newblock \bibinfo{title}{Consensus dynamics and opinion formation on
		hypergraphs}.
	\newblock In \emph{\bibinfo{booktitle}{Higher-Order Systems}},
	\bibinfo{pages}{347--376}, \doiprefix\url{10.1007/978-3-030-91374-8_14}
	(\bibinfo{publisher}{Springer}, \bibinfo{year}{2022}).
	
	\bibitem{skardal2022explosive}
	\bibinfo{author}{Skardal, P.~S.} \& \bibinfo{author}{Arenas, A.}
	\newblock \bibinfo{title}{Explosive synchronization and multistability in large
		systems of kuramoto oscillators with higher-order interactions}.
	\newblock In \emph{\bibinfo{booktitle}{Higher-Order Systems}},
	\bibinfo{pages}{217--232}, \doiprefix\url{10.1007/978-3-030-91374-8_8}
	(\bibinfo{publisher}{Springer}, \bibinfo{year}{2022}).
	
	\bibitem{millan2022geometry}
	\bibinfo{author}{Mill{\'a}n, A.~P.}, \bibinfo{author}{Restrepo, J.~G.},
	\bibinfo{author}{Torres, J.~J.} \& \bibinfo{author}{Bianconi, G.}
	\newblock \bibinfo{title}{Geometry, topology and simplicial synchronization}.
	\newblock In \emph{\bibinfo{booktitle}{Higher-Order Systems}},
	\bibinfo{pages}{269--299}, \doiprefix\url{10.1007/978-3-030-91374-8_11}
	(\bibinfo{publisher}{Springer}, \bibinfo{year}{2022}).
	
	\bibitem{Traulsen2006Evolution}
	\bibinfo{author}{Traulsen, A.} \& \bibinfo{author}{Nowak, M.~A.}
	\newblock \bibinfo{journal}{\bibinfo{title}{Evolution of cooperation by
			multilevel selection}}.
	\newblock {\emph{\JournalTitle{Proc. Natl. Acad. Sci. U.S.A.}}}
	\textbf{\bibinfo{volume}{103}}, \bibinfo{pages}{10952--10955},
	\doiprefix\url{10.1073/pnas.0602530103} (\bibinfo{year}{2006}).
	\newblock \eprint{https://www.pnas.org/doi/pdf/10.1073/pnas.0602530103}.
	
	\bibitem{perc2013evolutionary}
	\bibinfo{author}{Perc, M.}, \bibinfo{author}{G{\'o}mez-Gardenes, J.},
	\bibinfo{author}{Szolnoki, A.}, \bibinfo{author}{Flor{\'\i}a, L.~M.} \&
	\bibinfo{author}{Moreno, Y.}
	\newblock \bibinfo{journal}{\bibinfo{title}{Evolutionary dynamics of group
			interactions on structured populations: a review}}.
	\newblock {\emph{\JournalTitle{J. R. Soc. Interface}}}
	\textbf{\bibinfo{volume}{10}}, \bibinfo{pages}{20120997},
	\doiprefix\url{https://doi.org/10.1098/rsif.2012.0997}
	(\bibinfo{year}{2013}).
	
	\bibitem{schweitzer2022social}
	\bibinfo{author}{Schweitzer, F.} \& \bibinfo{author}{Andres, G.}
	\newblock \bibinfo{journal}{\bibinfo{title}{Social nucleation: Group formation
			as a phase transition}}.
	\newblock {\emph{\JournalTitle{Phys. Rev. E}}} \textbf{\bibinfo{volume}{105}},
	\bibinfo{pages}{044301}, \doiprefix\url{10.1103/PhysRevE.105.044301}
	(\bibinfo{year}{2022}).
	
	\bibitem{rosenthal2015revealing}
	\bibinfo{author}{Rosenthal, S.~B.}, \bibinfo{author}{Twomey, C.~R.},
	\bibinfo{author}{Hartnett, A.~T.}, \bibinfo{author}{Wu, H.~S.} \&
	\bibinfo{author}{Couzin, I.~D.}
	\newblock \bibinfo{journal}{\bibinfo{title}{Revealing the hidden networks of
			interaction in mobile animal groups allows prediction of complex behavioral
			contagion}}.
	\newblock {\emph{\JournalTitle{Proc. Natl. Acad. Sci. U.S.A.}}}
	\textbf{\bibinfo{volume}{112}}, \bibinfo{pages}{4690--4695},
	\doiprefix\url{https://doi.org/10.1073/pnas.1420068112}
	(\bibinfo{year}{2015}).
	
	\bibitem{iacopini2024not}
	\bibinfo{author}{Iacopini, I.}, \bibinfo{author}{Foote, J.~R.},
	\bibinfo{author}{Fefferman, N.~H.}, \bibinfo{author}{Derryberry, E.~P.} \&
	\bibinfo{author}{Silk, M.~J.}
	\newblock \bibinfo{journal}{\bibinfo{title}{Not your private
			t{\^e}te-{\`a}-t{\^e}te: leveraging the power of higher-order networks to
			study animal communication}}.
	\newblock {\emph{\JournalTitle{Philos. Trans. R. Soc. B: Biol. Sci.}}}
	\textbf{\bibinfo{volume}{379}}, \bibinfo{pages}{20230190},
	\doiprefix\url{https://doi.org/10.1098/rstb.2023.0190}
	(\bibinfo{year}{2024}).
	
	\bibitem{Gelardi2020MeasuringSN}
	\bibinfo{author}{Gelardi, V.}, \bibinfo{author}{Godard, J.},
	\bibinfo{author}{Paleressompoulle, D.}, \bibinfo{author}{Claidi{\`e}re, N.}
	\& \bibinfo{author}{Barrat, A.}
	\newblock \bibinfo{journal}{\bibinfo{title}{Measuring social networks in
			primates: wearable sensors versus direct observations}}.
	\newblock {\emph{\JournalTitle{Proc. R. Soc A}}}
	\textbf{\bibinfo{volume}{476}}, \bibinfo{pages}{20190737},
	\doiprefix\url{https://doi.org/10.1098/rspa.2019.0737}
	(\bibinfo{year}{2020}).
	
	\bibitem{flierl1999individuals}
	\bibinfo{author}{Flierl, G.}, \bibinfo{author}{Gr{\"u}nbaum, D.},
	\bibinfo{author}{Levin, S.} \& \bibinfo{author}{Olson, D.}
	\newblock \bibinfo{journal}{\bibinfo{title}{From individuals to aggregations:
			the interplay between behavior and physics}}.
	\newblock {\emph{\JournalTitle{J. Theor. Biol.}}}
	\textbf{\bibinfo{volume}{196}}, \bibinfo{pages}{397--454},
	\doiprefix\url{https://doi.org/10.1006/jtbi.1998.0842}
	(\bibinfo{year}{1999}).
	
	\bibitem{conradt2000activity}
	\bibinfo{author}{Conradt, L.} \& \bibinfo{author}{Roper, T.~J.}
	\newblock \bibinfo{journal}{\bibinfo{title}{Activity synchrony and social
			cohesion: a fission-fusion model}}.
	\newblock {\emph{\JournalTitle{Proc. Royal Soc. B}}}
	\textbf{\bibinfo{volume}{267}}, \bibinfo{pages}{2213--2218},
	\doiprefix\url{https://doi.org/10.1098/rspb.2000.1271}
	(\bibinfo{year}{2000}).
	
	\bibitem{wittemyer2005socioecology}
	\bibinfo{author}{Wittemyer, G.}, \bibinfo{author}{Douglas-Hamilton, I.} \&
	\bibinfo{author}{Getz, W.~M.}
	\newblock \bibinfo{journal}{\bibinfo{title}{The socioecology of elephants:
			analysis of the processes creating multitiered social structures}}.
	\newblock {\emph{\JournalTitle{Anim. Behav.}}} \textbf{\bibinfo{volume}{69}},
	\bibinfo{pages}{1357--1371},
	\doiprefix\url{https://doi.org/10.1016/j.anbehav.2004.08.018}
	(\bibinfo{year}{2005}).
	
	\bibitem{archie2006ties}
	\bibinfo{author}{Archie, E.~A.}, \bibinfo{author}{Moss, C.~J.} \&
	\bibinfo{author}{Alberts, S.~C.}
	\newblock \bibinfo{journal}{\bibinfo{title}{The ties that bind: genetic
			relatedness predicts the fission and fusion of social groups in wild african
			elephants}}.
	\newblock {\emph{\JournalTitle{Proc. Royal Soc. B}}}
	\textbf{\bibinfo{volume}{273}}, \bibinfo{pages}{513--522},
	\doiprefix\url{https://doi.org/10.1098/rspb.2005.3361}
	(\bibinfo{year}{2006}).
	
	\bibitem{gavrilets2015collective}
	\bibinfo{author}{Gavrilets, S.}
	\newblock \bibinfo{journal}{\bibinfo{title}{Collective action problem in
			heterogeneous groups}}.
	\newblock {\emph{\JournalTitle{Philos. Trans. R. Soc. B}}}
	\textbf{\bibinfo{volume}{370}}, \bibinfo{pages}{20150016},
	\doiprefix\url{https://doi.org/10.1098/rstb.2015.0016}
	(\bibinfo{year}{2015}).
	
	\bibitem{alvarez2021evolutionary}
	\bibinfo{author}{Alvarez-Rodriguez, U.} \emph{et~al.}
	\newblock \bibinfo{journal}{\bibinfo{title}{Evolutionary dynamics of
			higher-order interactions in social networks}}.
	\newblock {\emph{\JournalTitle{Nat. Hum. Behav.}}}
	\textbf{\bibinfo{volume}{5}}, \bibinfo{pages}{586--595},
	\doiprefix\url{https://doi.org/10.1038/s41562-020-01024-1}
	(\bibinfo{year}{2021}).
	
	\bibitem{young2021hypergraph}
	\bibinfo{author}{Young, J.-G.}, \bibinfo{author}{Petri, G.} \&
	\bibinfo{author}{Peixoto, T.~P.}
	\newblock \bibinfo{journal}{\bibinfo{title}{Hypergraph reconstruction from
			network data}}.
	\newblock {\emph{\JournalTitle{Commun. Phys.}}} \textbf{\bibinfo{volume}{4}},
	\bibinfo{pages}{135},
	\doiprefix\url{https://doi.org/10.1038/s42005-021-00637-w}
	(\bibinfo{year}{2021}).
	
	\bibitem{musciotto2021detecting}
	\bibinfo{author}{Musciotto, F.}, \bibinfo{author}{Battiston, F.} \&
	\bibinfo{author}{Mantegna, R.~N.}
	\newblock \bibinfo{journal}{\bibinfo{title}{Detecting informative higher-order
			interactions in statistically validated hypergraphs}}.
	\newblock {\emph{\JournalTitle{Commun. Phys.}}} \textbf{\bibinfo{volume}{4}},
	\bibinfo{pages}{218},
	\doiprefix\url{https://doi.org/10.1038/s42005-021-00710-4}
	(\bibinfo{year}{2021}).
	
	\bibitem{alvarez2023inference}
	\bibinfo{author}{Alvarez-Rodriguez, U.}, \bibinfo{author}{Petrovi{\'c}, L.~V.}
	\& \bibinfo{author}{Scholtes, I.}
	\newblock \bibinfo{journal}{\bibinfo{title}{Inference of time-ordered multibody
			interactions}}.
	\newblock {\emph{\JournalTitle{Phys. Rev. E}}} \textbf{\bibinfo{volume}{108}},
	\bibinfo{pages}{034312}, \doiprefix\url{10.1103/PhysRevE.108.034312}
	(\bibinfo{year}{2023}).
	
	\bibitem{DTU_corse_base}
	\bibinfo{author}{of~Denmark, T.~U.}
	\newblock \bibinfo{title}{Course base}.
	\newblock
	\bibinfo{howpublished}{\url{https://www.dtu.dk/english/education/course-base}}
	(\bibinfo{year}{2022}).
	
	\bibitem{sekara2014strength}
	\bibinfo{author}{Sekara, V.} \& \bibinfo{author}{Lehmann, S.}
	\newblock \bibinfo{journal}{\bibinfo{title}{The strength of friendship ties in
			proximity sensor data}}.
	\newblock {\emph{\JournalTitle{PLoS One}}} \textbf{\bibinfo{volume}{9}},
	\bibinfo{pages}{e100915},
	\doiprefix\url{https://doi.org/10.1371/journal.pone.0100915}
	(\bibinfo{year}{2014}).
	
	\bibitem{hall1990hidden}
	\bibinfo{author}{Hall, E.}
	\newblock \emph{\bibinfo{title}{The Hidden Dimension}}
	(\bibinfo{publisher}{Anchor Books}, \bibinfo{year}{1990}).
	
	\bibitem{reis1991studying}
	\bibinfo{author}{Reis, H.~T.} \& \bibinfo{author}{Wheeler, L.}
	\newblock \bibinfo{journal}{\bibinfo{title}{Studying social interaction with
			the rochester interaction record}}.
	\newblock {\emph{\JournalTitle{Adv. Exp. Soc. Psychol.}}}
	\textbf{\bibinfo{volume}{24}}, \bibinfo{pages}{269--318},
	\doiprefix\url{https://doi.org/10.1016/S0065-2601(08)60332-9}
	(\bibinfo{year}{1991}).
	
	\bibitem{Landry2023}
	\bibinfo{author}{Landry, N.~W.} \emph{et~al.}
	\newblock \bibinfo{journal}{\bibinfo{title}{Xgi: A python package for
			higher-order interaction networks}}.
	\newblock {\emph{\JournalTitle{J. Open Source Softw.}}}
	\textbf{\bibinfo{volume}{8}}, \bibinfo{pages}{5162},
	\doiprefix\url{10.21105/joss.05162} (\bibinfo{year}{2023}).
	
	\bibitem{iacopini2024temporal_code}
	\bibinfo{author}{Iacopini, I.}, \bibinfo{author}{Karsai, M.} \&
	\bibinfo{author}{Barrat, A.}
	\newblock \bibinfo{journal}{\bibinfo{title}{The temporal dynamics of group
			interactions in higher-order social networks}}.
	\newblock {\emph{\JournalTitle{iaciac/temporal-group-interactions}}}
	\doiprefix\url{10.5281/zenodo.12698353} (\bibinfo{year}{2024}).
	
	\bibitem{sociopatterns}
	\bibinfo{title}{{SocioPatterns Collaboration}}.
	\newblock \bibinfo{howpublished}{\url{http://www.sociopatterns.org/}}.
	
\end{thebibliography}


\pagebreak[2]
\clearpage
\newpage
\startsupplement
\renewcommand{\theequation}{Eq. S\arabic{equation}}
\let\oldsection\section
\renewcommand{\section}[1]{\oldsection{Supplementary Note~\thesection \ -- \ #1}} 
\tableofcontents
\listoffigures
\listoftables

\newpage

\section{GESIS data sets on interactions at conferences}\label{SI:sec:Confs}

We use public data collected by the SocioPatterns collaboration~\cite{sociopatterns} during four different scientific conferences organised by GESIS, the Leibniz Institute for Social Sciences, Germany: the  3rd  GESIS  Computational  Social  Science  Winter  Symposium in 2016 (WS16), the  International  Conference  on  Computational  Social  Science in 2017 (ICCSS17), the Eurosymposium on Computational Social Science in 2018 (ECSS18), and the 41st European Conference on Information Retrieval in 2019 (ECIR19). These conferences were strongly interdisciplinary, covering a wide range of topics from computer to social and natural sciences. The data consists in face-to-face interactions among participants of the different conferences as measured by wearable RFID sensors with a temporal resolution of 20 seconds. The details of the data collection, including the collection technique and the analysis of the interactions, are presented in the original Ref.~\cite{genois2023combining}.
The raw data for each conference contains 153,371 (WS16), 229,536 (ICCSS17), 96,362 (ECSS18), and 132,949 (ECIR19) entries, respectively. We aggregate these records, as per the University data set, using a temporal window of 5 minutes, ending up with 38,832 (WS16), 58,745 (ICCSS17), 24,637 (ECSS18), and 34,217 (ECIR19) entries. We then add isolated nodes to the data. In particular, for each conference, we consider all the nodes that ever appear and add a record of the node being isolated for each time step after their first appearance. At this point the data are composed of 46,915 (WS16), 98,124 (ICCSS17), 35,124 (ECSS18), and 61,371 (ECIR19) entries.

\clearpage

\begin{table}
	\begin{center} 
		\begin{tabular}{| l | l | c |} 
			\hline
			Data set & Context & Unique membership (frac.)\\
			\hline
			University & in-class & 0.75\\
                University & out-of-class & 0.96\\
			University & weekend & 0.98\\
                \hline
                Preschool & in-class & 0.80\\
                Preschool & out-of-class & 0.94\\
                \hline
		\end{tabular}
		\caption[Fractions of single-membership interactions]{
  Fractions of single-membership interactions.
  }
		\label{SI:table:membership}
	\end{center}
\end{table}

\end{document}